\documentclass{article}

\usepackage{authblk}
\usepackage[utf8]{inputenc}
\usepackage{main}
\usepackage{microtype}
\usepackage{subcaption}
\usepackage{graphicx}
\usepackage{times}
\usepackage{latexsym}
\usepackage{amsmath}
\usepackage{float}
\usepackage{footnote}
\usepackage{enumitem}
\usepackage{bm}
\usepackage{arydshln}
\usepackage{booktabs}
\usepackage{multicol}
\usepackage{multirow}
\usepackage{color}
\usepackage{xcolor}     
\usepackage{colortbl}
\usepackage{bbding}
\usepackage{makecell}
\usepackage{mathtools}
\usepackage{imakeidx}
\usepackage{longtable}
\usepackage{wrapfig}
\usepackage{rotating}
\makeindex
\usepackage{arydshln}
\usepackage{lipsum}
\usepackage{natbib}
\usepackage[toc]{multitoc}
\usepackage[edges]{forest}
\usepackage[normalem]{ulem}
\definecolor{mydarkblue}{rgb}{0,0.08,0.45}
\usepackage[colorlinks=true,linkcolor=mydarkblue,citecolor=mydarkblue,filecolor=mydarkblue,urlcolor=mydarkblue]{hyperref}
\usepackage{CJKutf8}
\usepackage{awesomebox} 
\usepackage{bbding}
\usepackage[most]{tcolorbox}
\usepackage{booktabs}
\usepackage{geometry}
\definecolor{wkblue}{rgb}{0.2, 0.3, 0.6}
\definecolor{meta-color}{rgb}{0.5, 0.5, 0.5}
\usepackage{amsmath}
\usepackage{enumitem}
\usepackage{lscape} 
\usepackage{booktabs}
\usepackage{algorithm}      
\usepackage{algorithmicx}
\usepackage{algpseudocode}
\usepackage{tabularx,booktabs}
\usepackage{makecell}

\usepackage{amssymb}
\usepackage{amsfonts}

\usepackage[tikz]{bclogo}
\usepackage[framemethod=tikz]{mdframed}
\definecolor{bgblue}{RGB}{245,243,253}
\definecolor{ttblue}{RGB}{91,194,224}

\usepackage{fancyhdr} 
\usepackage{blindtext} 
\usepackage{makecell}

\usepackage{tikz}
\usepackage[edges]{forest}
\usepackage{microtype}
\usepackage{hyperref}
\usepackage{url}
\usepackage{booktabs}
\usepackage{bbding}
\usepackage{multirow}
\usepackage{booktabs}
\usepackage{float}
\usepackage{makecell}

\usepackage{times}
\usepackage{latexsym}
\usepackage{url}
\usepackage{forest}
\usepackage{setspace}
\usepackage{placeins}
\usepackage{booktabs}
\usepackage{multirow}
\usepackage{tabularx}
\usepackage{tabulary}
\usepackage{graphicx}
\usepackage{adjustbox}
\usepackage{colortbl}
\usepackage{xcolor}
\usepackage{enumitem}
\usepackage{amsfonts}
\usepackage{bbding}
\usepackage{wrapfig}
\usepackage{amsmath}
\usepackage{hyperref}
\usepackage{pifont}
\usetikzlibrary{trees,positioning,shapes,shadows,arrows.meta}
\usepackage{chemformula}

\usepackage[edges]{forest}
\usepackage[most]{tcolorbox}
\usepackage{lipsum} 
\usepackage{fontawesome5}  

\title{Beyond Message Passing: A Semantic View of Agent Communication Protocols}
\author{
Dun Yuan,
Fuyuan Lyu\thanks{Co-first author.},
Ye Yuan,
Weixu Zhang,
Bowei He,
Jiayi Geng,
Linfeng Du,
Zipeng Sun,
Yankai Chen,
Changjiang Han,
Jikun Kang,
Xi Chen,
Haolun Wu,
Xue Liu
}

\date{September 2025}

\begin{document}

\maketitle

\begin{abstract}
Agent communication protocols are becoming critical infrastructure for large language model (LLM) systems that must use tools, coordinate with other agents, and operate across heterogeneous environments. This survey presents a human-inspired perspective on this emerging landscape by organizing agent communication into three layers: communication, syntactic, and semantic. Under this framework, we systematically analyze 18 representative protocols and compare how they support reliable transport, structured interaction, and meaning-level coordination.
Our analysis shows a clear imbalance in current protocol design. Most protocols provide increasingly mature support for transport, streaming, schema definition, and lifecycle management, but offer limited protocol-level mechanisms for clarification, context alignment, and verification. As a result, semantic responsibilities are often pushed into prompts, wrappers, or application-specific orchestration logic, creating hidden interoperability and maintenance costs. To make this gap actionable, we further identify major forms of technical debt in today’s protocol ecosystem and distill practical guidance for selecting protocols under different deployment settings. We conclude by outlining a research agenda for interoperable, secure, and semantically robust agent ecosystems that move beyond message passing toward shared understanding.
\end{abstract}

\section{Introduction}


With the rapid progress of large language models (LLMs), AI-powered agents have become increasingly popular and widely deployed across domains such as customer service, content creation, healthcare, and more. To unlock their full potential, agents must exchange information with one another~\cite{hong2023metagpt,wu2024autogen} and with external resources such as data stores and tools~\cite{schick2023toolformer,yao2022react}.

As agent communication shares similarities with human beings, it is helpful to recall how humans communicate. At a high level, effective human communication has three aspects. First, there must be a channel so both sides can speak and hear each other through a given medium, such as waves in the air or digits in telephones. Second, there must be a shared syntax, such as English or Chinese, so that messages are syntactically understandable. Third and most crucial, there must be a process for aligning meaning. Humans do not merely parse syntax; instead, they actively collaborate to achieve mutual understanding. Faced with an ambiguous instruction like “Book a flight to Springfield,” a human travel agent asks, “Which Springfield, the one in Illinois or Massachusetts?” Clarifying questions, confirmations, and repairs are the mechanisms by which humans bridge what is said and what is meant.

As agents become increasingly powerful, they take on complex, diverse tasks for which domain-specific designs are insufficient. The community is aware that standardized ways for agents to connect are needed~\cite{anthropic_model_2024}. Various agent communication protocols have emerged. These protocols function as the "messenger carrier" between AI agents, providing a common language and a standardized framework for structured interaction among LLM-powered agents. Standards such as the Model Context Protocol (MCP)~\cite{anthropic_model_2024}, the Agent Communication Protocol (ACP)~\cite{agentunion_acp}, and the Agent-to-Agent (A2A)~\cite{google_a2a_2025} Protocol aim to transform disconnected collections of individual agents into cohesive intelligent systems capable of tackling complex real-world challenges. They are the essential infrastructure for creating an interconnected "Internet of Agents"~\cite{internet_of_agents}. Existing surveys \citep{kong2025survey,yang2025survey,ehtesham2025survey,deng2025ai} provide comprehensive summaries and analyses of ongoing trends in this field.


Inspired by the human communication process, we propose a three-layer framework for evaluating and analyzing existing agent communication protocols. Our taxonomy comprises three distinct layers: \textbf{communication}, \textbf{syntactic}, and \textbf{semantic}, with formal definitions provided in Section~\ref{sec:taxonomy}.

\begin{figure*}[t]
    \centering
    \includegraphics[width=0.9\textwidth]{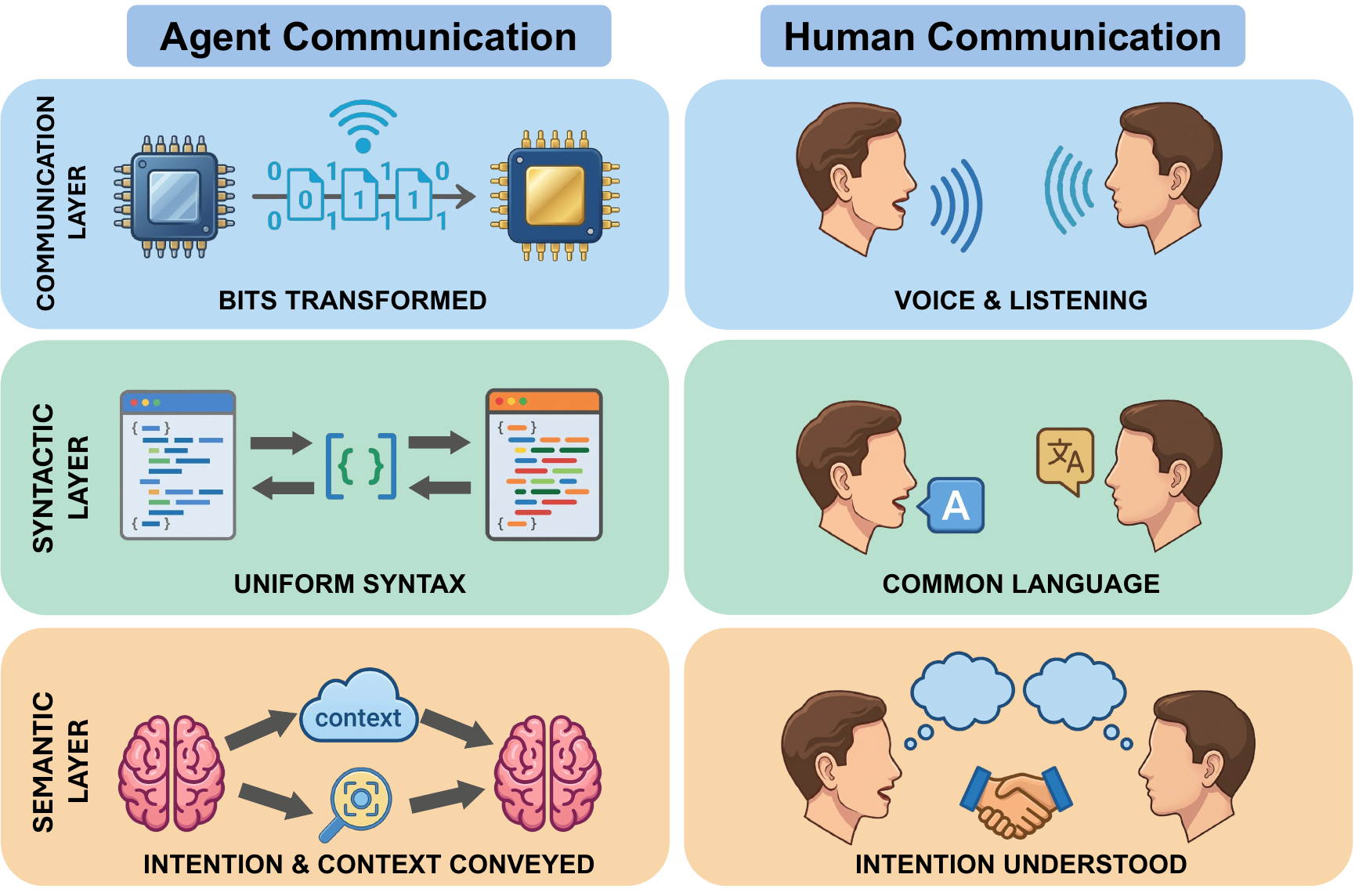}
    \caption{Human communication and agent communication can be compared through the same three-layer lens: communication, syntactic, and semantic alignment.}
    \label{fig:agents-vs-human}
\end{figure*}

In this survey, we systematically analyze 18 agent communication protocols through the lens of human communication. Our analysis reveals that most existing protocols prioritize the first two layers—establishing communication channels and defining syntactic standards—while providing limited explicit support for semantic alignment. In essence, contemporary agents can reliably transmit and parse messages, yet they lack built-in mechanisms for clarification, confirmation, and repair that would render their interactions robust against ambiguity.

While previous surveys~\cite{yang2025survey,kong2025survey,ehtesham2025survey,ray2025review,deng2025ai} on AI agent communication protocols have primarily focused on operational mechanics, architectural comparisons, and security vulnerabilities, our survey distinguishes itself by analyzing these systems through the lens of human communication theory. For example, recent works such as \citet{kong2025survey} and \citet{deng2025ai} center heavily on mapping out security risks, attack surfaces, and defense countermeasures across different communication stages (user-agent, agent-agent, and agent-environment). Other surveys, such as \citet{yang2025survey} and \citet{ray2025review}, classify protocols based on context-orientation and evaluate technical performance metrics like scalability, latency, and privacy. Similarly, \citet{ehtesham2025survey} provides an architectural comparison of a few specific interoperability standards to propose a phased adoption roadmap. In contrast to existing surveys, this paper introduces a unique framework inspired by human communication, dividing protocols into communication, syntactic, and semantic layers. We summarize our key contributions as follows:
\begin{itemize}
    \item We establish a formal correspondence between human communication theory and agent communication protocols, proposing a three-layer taxonomy that enables systematic evaluation of protocol completeness and illuminates pathways for future research.
    \item We conduct a comprehensive analysis of 18 existing protocols using our taxonomy, providing actionable insights that help practitioners select protocols aligned with their specific requirements.
    \item We identify and analyze technical debt accumulated during the rapid development of agent communication protocols, offering recommendations for sustainable architectural evolution.
\end{itemize}

The remainder of this paper is structured as follows. Section~\ref{sec:preliminary} establishes foundational concepts covering AI agents, communication protocols, and human communication theory. Section~\ref{sec:taxonomy} presents our three-layer taxonomy in detail. Section~\ref{sec:protocol} applies this taxonomy to analyze 18 representative protocols. Section~\ref{sec:techdebt} examines technical debt accumulated during the rapid evolution of agent communication systems. Finally, Section~\ref{sec:go-to-option} offers evidence-based recommendations to guide protocol selection and development according to specific use cases.

\section{Preliminary}
\label{sec:preliminary}

\subsection{Definition of AI Agents}
AI agents have undergone substantial evolution over the past three decades, transitioning from symbolic rule-based systems to modern foundation-model–driven autonomous actors. At a high level, an \emph{AI agent} is a computational entity capable of perceiving its environment, making decisions, and executing actions to pursue designated objectives \citep{franklin1996agent,jennings2000agent,wooldridge2009introduction}. This perception--decision--action loop underlies diverse agent paradigms including classical multi-agent systems (MAS), robotic control systems, web-service agents, and the most recent generation of large-model–powered agents.

\paragraph{From Classical MAS to Modern Agentic AI.}
Early agent research in the 1990s–2000s treated agents as autonomous problem solvers operating within simulations or distributed service frameworks.  
Protocols such as \textbf{KQML} and \textbf{FIPA-ACL} formalized communication through symbolic performatives and logical operators, enabling agents to exchange commitments, beliefs, intentions, and queries \citep{kqml,fipa2002communicative}.  
Although these systems offered rich semantic formalisms, they were tightly coupled to handcrafted ontologies and often too rigid or brittle for deployment in open-ended environments.

From 2010–2020, research diversified into robotics, game agents, and service orchestration, yet progress on communication standards stagnated.  
Frameworks such as the Contract Net Protocol \citep{smith1988contract}, negotiation protocols for robotic coordination, and semantic-web agents (e.g., OWL-S, WS-Agents \citep{greenwood2005semantic,martin2004owl,curbera2001web}) contributed important foundations but faced adoption challenges due to complexity, ontology maintenance costs, and limited compatibility with modern API ecosystems \citep{milojicic1998masif,hohpe2004enterprise}.  
These systems retained expressive formal semantics but lacked the flexibility required for broad interoperability.

\paragraph{Rise of Modern AI Agents.}
The breakthrough of large language models (LLMs) and multimodal foundation models has redefined the notion of an agent.  
Modern AI agents are no longer restricted to symbolic reasoning or fixed action schemas; instead, they integrate perception, natural-language reasoning, memory, tool invocation, and adaptive planning \citep{openai_function_calling}.  
This enables agents to decompose tasks, ground decisions in external information, interact with heterogeneous APIs, and collaborate with humans or other agents in natural language \citep{yang2025survey,kong2025survey}.

Among various forms of AI agents---robotic agents, software assistants, API-driven services, autonomous decision-making modules---\emph{LLM-driven agents} have emerged as the dominant paradigm due to their broad generalization capabilities and flexible linguistic interface.  
LLMs allow agents to operate over unstructured instructions, generate executable plans, and coordinate through language-like messages rather than brittle predefined schemas \citep{autogpt,crewai,ag2}.  
This shift has revitalized the long-standing vision of an ``\textbf{Internet of Agents}'' \citep{internet_of_agents}, where heterogeneous AI systems communicate and collaborate as composable, distributed entities.

\paragraph{Motivation for Protocol Standardization.}

As AI agents become increasingly autonomous and interconnected, especially in tool-rich and networked environments, their actions can directly affect external systems and real-world processes. This raises the need for communication mechanisms that are not only standardized and reliable, but also interpretable across heterogeneous agents and services. Without such protocols, developers must rely on ad-hoc wrappers, prompt conventions, or proprietary interfaces, which may work locally but scale poorly and often fail to preserve consistent behavior across deployments.

More importantly, the challenge is no longer merely enabling agents to exchange messages. Modern agents can already call tools, invoke APIs, and coordinate through structured interfaces \citep{chase2022langchain,liu2023llamaindex,openai_plugins}. The deeper challenge is ensuring that they interpret requests, constraints, and outcomes in compatible ways. This motivates the development of modern agent communication protocols, which we introduce next.


\subsection{Agent Communication Protocols}
As agents transition from isolated systems to components of larger cooperative ecosystems, they require structured frameworks for exchanging information and coordinating actions. An \emph{agent communication protocol} (ACP) defines the rules, message structures, and interaction patterns that make such coordination possible. Much like Internet protocols standardize communication among computers, ACPs provide the foundational interface through which AI agents discover capabilities, issue requests, share results, and coordinate behavior \citep{milev2025toolfuzz,sheriff2024metadata_manifest}.

However, standardizing communication is not the same as standardizing understanding. In practice, a protocol may ensure that messages are delivered correctly and conform to a shared schema, while still leaving substantial ambiguity about how requests should be interpreted or verified. This distinction between successful message exchange and successful meaning alignment is central to our analysis.

\paragraph{Historical Context.}
Classical protocols such as KQML and FIPA-ACL emphasized symbolic performatives and logic-based semantics \citep{kqml,fipa2002communicative}.  
However, their reliance on formal ontologies made them challenging to scale and cumbersome to integrate with web-era APIs or dynamic environments.  
Similarly, semantic-web–oriented frameworks (e.g., OWL-S, WS-Agents \citep{martin2004owl,greenwood2005semantic}) provided expressive structuring mechanisms but saw limited real-world adoption due to brittleness and engineering overhead.

In contrast, modern agent ecosystems require lightweight, modular, cross-platform protocols that integrate smoothly with existing software stacks and support dynamic discovery of agent capabilities.

\paragraph{Modern Agent Protocols (2023–2025).}
The emergence of LLM-based agents triggered a renewed wave of protocol innovation.  
Well-known contemporary frameworks include:

\begin{itemize}[leftmargin=1.3em]
    \item \textbf{Model Context Protocol (MCP)} by Anthropic \citep{anthropic_model_2024} — defines standardized transport channels (stdio, WebSocket, HTTP), capability registration, and structured JSON-based tool invocation.
    \item \textbf{Agent-to-Agent (A2A)} by Google \citep{google_a2a_2025} — supports cross-platform agent messaging, capability negotiation, and secure agent-level identity management.
    \item \textbf{Agent Communication Protocol (ACP)} by AgentUnion \citep{agentunion_acp} — focuses on uniform message schemas, extensible agent actions, and multi-agent coordination.
    \item Community-driven standards such as \textbf{ANP}, \textbf{Coral}, and \textbf{LMOS} \citep{gaowei_chang_anp_2024,coral,eclipse_language_2025} — which explore decentralized interoperability, shared ontologies, and semantic grounding for agent networks.
\end{itemize}

Together, these protocols aim to transform diverse agents and tools into interoperable components within a broader agentic ecosystem.

\paragraph{Why Communication Protocols Matter.}



Modern AI agents---particularly LLM-driven ones---can act on external resources such as APIs, software tools, databases, and robots, which raises the stakes for correctness, safety, and coordination. Communication protocols therefore serve as core infrastructure for three distinct functions:

\begin{itemize}[leftmargin=1.3em]
    \item \textbf{Reliable communication}: standardized transport, message encoding, and error handling.
    \item \textbf{Shared syntax}: well-defined schemas for tool calls, agent actions, and capability descriptions.
    \item \textbf{Semantic alignment}: protocol-level support for clarifying intent, grounding context, and verifying that an action preserves the intended meaning.
\end{itemize}

Existing protocols have made visible progress on the first two functions, but the third remains comparatively underdeveloped. As a result, many systems can exchange messages correctly without guaranteeing that collaborating agents interpret those messages in the same way. This gap motivates the human-inspired taxonomy introduced in Section~\ref{sec:taxonomy}.

\subsection{Human Communication and Interdisciplinary Motivation}

Communication has long been studied across linguistics, cognitive science, pragmatics, and information theory, offering conceptual tools that can inform the design of robust agent communication protocols.  
While engineering perspectives often focus on message encoding and transport, human communication research emphasizes that meaning is not merely transmitted but \emph{co-constructed}.  
Bringing these perspectives together provides a principled foundation for analyzing the strengths and limitations of modern agent protocols.

\paragraph{Foundations from Information Theory.}
Classical models such as the Shannon–Weaver communication framework conceptualize communication as the reliable transmission of signals through a channel \citep{shannon1948}.  
This view aligns with the \textbf{communication layer} in agent systems: ensuring that messages are delivered correctly, efficiently, and with minimal noise.  
Early MAS protocols inherited this framing, prioritizing structured message passing but offering limited support for contextual interpretation.

\paragraph{Linguistic and Pragmatic Perspectives.}
Human communication relies on shared conventions that go beyond channel transmission.  
Linguistics distinguishes between the syntactic structure of messages and the mechanisms for mapping those structures onto meaning.  
Pragmatics further studies how speakers use context and inference to interpret ambiguous or underspecified utterances.  

Influential frameworks include:
\begin{itemize}[leftmargin=1.3em]
    \item \textbf{Grice's Cooperative Principle} \citep{grice1975} — speakers follow maxims of quality, quantity, relation, and manner to collaboratively construct meaning.
    \item \textbf{Clark's Theory of Common Ground} \citep{clark1996} — communication is a joint activity requiring mutual belief about shared knowledge, intentions, and assumptions.
    \item \textbf{Relevance Theory} \citep{sperber1995} — interpretation is guided by the search for the most contextually relevant meaning with minimal cognitive effort.
\end{itemize}

These frameworks highlight a crucial insight: humans seldom interpret messages literally; instead, they actively infer the speaker's intended meaning using clarification, confirmation, repair, and contextual grounding.

\paragraph{Gap Between Human and Agent Communication.}
Most current agent communication protocols implement only the first two layers of human communication:
\begin{itemize}[leftmargin=1.3em]
    \item \textbf{Communication Layer}: establishing transport channels (e.g., HTTP, WebSocket, stdio) and ensuring message reliability.
    \item \textbf{Syntactic Layer}: defining schemas, typed message formats, and interfaces for tool invocation or capability registration.
\end{itemize}

However, they provide limited support for the \textbf{semantic layer}---the layer responsible for aligning meaning, resolving ambiguity, verifying intent, and establishing shared context.  
Protocols like MCP and A2A ensure that a JSON message is well-formed, but do not determine whether two agents interpret a request the same way, nor do they specify how agents should negotiate meaning when ambiguity arises.  

For example, if one agent receives the instruction ``Book a flight to Springfield,'' current protocols transport the message and parse it correctly, but lack built-in mechanisms for semantic repair (e.g., ``Which Springfield do you mean?'') or for maintaining persistent shared context analogous to human common ground.

\paragraph{Motivating a Human-Inspired Taxonomy.}

These interdisciplinary perspectives suggest that effective agent communication depends on three interdependent layers:
\begin{enumerate}[leftmargin=1.3em,itemsep=0.3em]
    \item \textbf{Communication Layer} --- ensures reliable transmission of signals (``Can we send and receive messages reliably?'').
    \item \textbf{Syntactic Layer} --- ensures structural compatibility and shared message formats (``Are we speaking in a mutually interpretable language?'').
    \item \textbf{Semantic Layer} --- ensures intent alignment, ambiguity resolution, and shared understanding (``Do we mean the same thing?'').
\end{enumerate}

This three-layer view also clarifies the historical tradeoff in agent communication research. Classical MAS protocols often pursued rich symbolic semantics at the cost of flexibility and deployability, whereas many modern protocols prioritize practical integration while leaving meaning-level coordination weakly specified. Our central argument is that robust agent ecosystems require all three layers to work together. This observation motivates the taxonomy in Section~\ref{sec:taxonomy}, which evaluates existing protocols through these dimensions and highlights opportunities to improve the semantic robustness of agent communication.

\section{Three-Layer Taxonomy} 
\label{sec:taxonomy}


\begin{figure*}[ht]
    \centering
    \includegraphics[width=0.92\textwidth]{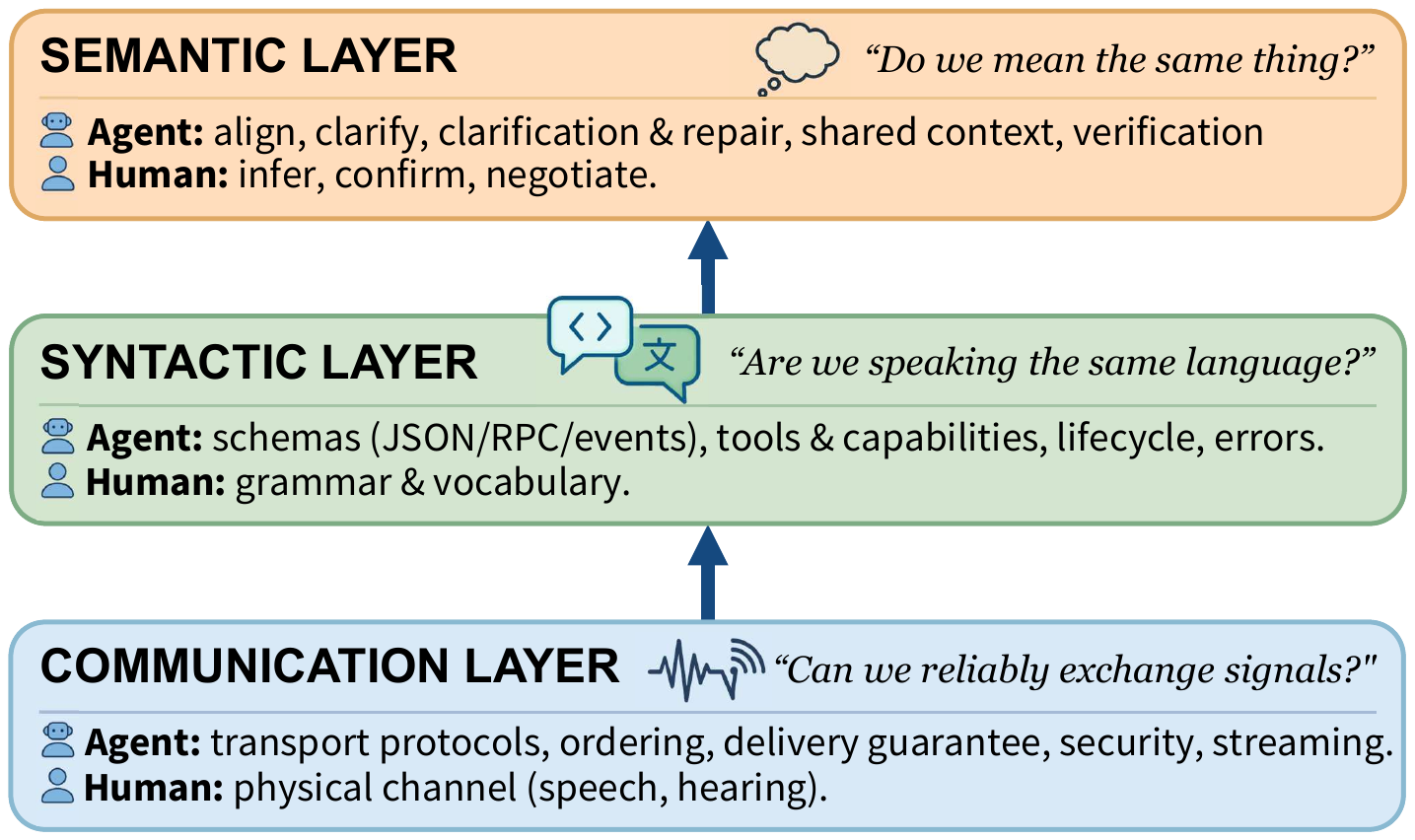}
    \caption{Overview of the proposed three-layer taxonomy for agent communication protocols, spanning communication, syntactic, and semantic concerns.}
    \label{fig:three-layer-taxonomy}
\end{figure*}

\subsection{Communication Layer}
The Communication Layer provides the physiological foundation for agent interaction, much as the sensory and vocal systems do in human communication. Before meaning or understanding can emerge, signals must be physically transmitted and perceived. The Communication Layer ensures that agents can “hear” and “speak” to one another through reliable transfer of bytes, events, and artifacts without assuming anything about their meaning. It provides link reliability, routing, congestion control, and multiplexing, and exposes only a small, stable transport envelope (e.g., protocol ID, stream ID, sequence, timestamp, headers) plus clear delivery states (open, chunk, end, error) and backpressure signals. 


\paragraph{Common types of communication.}
To support diverse tasks and interaction patterns, this layer standardizes three delivery modes—text streams, tool calls, and file transfers—each with optional chunking and streaming:
\begin{enumerate}
    \item \textbf{Text streams.}
  Text is delivered progressively, so interactive exchanges render within tens of milliseconds instead of waiting for completion.
  \item \textbf{Tool calls.}
 Tool calls travel as typed requests and responses on the same ordered stream, preserving sequencing and making backpressure visible to both sides.
  \item \textbf{File transfers.}
  Large files are split into verifiable chunks with sequence numbers and integrity digests, allowing receivers to validate and resume transfers mid-way.
\end{enumerate}

\paragraph{Reliability.}
We define reliability as the transport’s ability to deliver every message or chunk in accordance with an explicit per-stream contract for delivery and ordering, despite transient failures, variable latency, or endpoint restarts. The contract specifies the delivery guarantees—ordering (in-order within a stream), delivery mode (at-least-once by default; at-most-once via deduplication and idempotency keys; “effectively once” when paired with idempotent operations), and completion visibility (acknowledged commit points and clear terminal vs.\ retryable outcomes).

\paragraph{Communication security.} This layer hardens the byte stream and its control channels, committing to confidentiality, integrity and authenticity, freshness and replay resistance, error detection/loss recovery, and safe secret handling so upper layers can build semantics, policy, and authorization on a trustworthy transport. 
Confidentiality and integrity in transit rely on modern TLS (and mutual TLS for service-to-service calls). Where intermediaries exist, per-frame message authentication codes or signatures can extend integrity beyond hop-by-hop encryption. Freshness and anti-replay protections use nonces and monotonically increasing sequence numbers bound to the stream and the caller identity.

\paragraph{Error detection.}
The aspect of error detection addresses non-adversarial faults in transit, looking for accidental corruption and loss while restoring a clean, ordered byte stream. Mechanisms include per-chunk checksums and validated reassembly for large file transfers, bounded retries with exponential backoff, resume tokens or byte/frame offsets for interrupted streams, optional forward error correction on lossy links, and liveness via heartbeats/timeouts. These guarantees export clear commit points and retryable vs.\ terminal outcomes to upper layers.



\subsection{Syntactic Layer}
The Syntactic Layer corresponds to the linguistic stage of human communication, where shared grammar and vocabulary enable structured expression. Just as humans rely on syntax to organize words into meaningful sentences, agents depend on defined schemas to structure their exchanges. The Syntactic Layer specifies how agents introduce themselves, describe their capabilities, and format their requests and responses. It defines the organization of inputs, outputs, and errors, regardless of how messages are transmitted. While the Communication Layer ensures that messages are delivered, the Syntactic Layer ensures that both sides use a common language, allowing for consistent interpretation and coordination.


\paragraph{Standardization.} Rather than hard-coding a single universal message shape for all agents, the Syntactic Layer requires each agent to publish precise schemas and capability descriptors as its standards. For instance, MCP \cite{anthropic_model_2024} builds on JSON-RPC 2.0 with a compact envelope and standardized error representations, so clients can parse and recover uniformly. A2A likewise defines a uniform message schema with optional attachments and an explicit task lifecycle with identifiers and states, enabling consistent coordination across vendors. ANP \cite{gaowei_chang_anp_2024}’s Agent Description (JSON-LD) similarly advertises callable actions with explicit input/output schemas and human-authorization policies. This approach allows diverse tasks and domains to coexist without forcing lowest-common-denominator payloads, while still supporting discovery, validation, and tool usage.


\paragraph{Error handling.} The Syntactic Layer standardizes stable error codes, human-readable messages, retry advice, and optional remediation hints linked to the schema that failed validation. This allows clients to present valuable feedback, implement a backoff strategy, or repair a request without having to guess. For example, MCP \cite{anthropic_model_2024} includes standardized error representations, status codes, and schema-based validation, enabling agents to reason about failures and implement robust recovery strategies. In Agora \cite{samuele_marro_scalable_2024}, when validation fails, the responder returns a structured error object with retry advice and diagnostic context, allowing the caller to correct the payload and retry or fall back to a fresh negotiation. The same applies to authentication and authorization signals: the contract indicates what credentials are required and how consent or scopes are conveyed, while the actual credential transport remains the responsibility of the Communication Layer.

\medskip
The Syntactic Layer is the middle stratum of the three-layer framework. Its defining characteristic is a transport-agnostic, typed message contract—schemas that turn delivered bytes into interpretable actions. It depends on the Communication Layer below for delivery, ordering, and security, and in turn enables the Semantic Layer to plan, verify, and enforce policies against machine-checkable inputs and outputs, as well as discoverable capabilities.

\subsection{Semantic Layer}
This layer represents the cognitive level of communication, where understanding and shared meaning emerge. Just as humans interpret sentences to grasp intent and context, agents rely on the Semantic Layer to align interpretation of what messages truly represent. While the Communication Layer ensures that information is transmitted and the Syntactic Layer organizes its form, the Semantic Layer gives that information purpose by defining intents, entities, and their relationships. It ensures that agents interpret the same request consistently, that actions are meaningful and safe, and that others can seamlessly reuse results. Achieving this alignment requires shared vocabularies, stable concept identifiers, and interpretable constraints that preserve meaning across time and implementations.

\paragraph{Intended outcome.}
The intended outcome represents a clear, machine-checkable task specification: the goal state, the parts of the system that may be read or written, and what constitutes success. 
Each operation states its preconditions and postconditions, the effects it guarantees, and any required evidence.
Data are typed beyond surface syntax—units, enumerations, identifiers, time stamps with time zones, and, when needed, uncertainty—so checkers can validate meaning, not just shape. 
Defined this way, plans expose explicit goals, subgoals, and dependencies, and cross-agent handoffs align on goals and guarantees rather than on field reshaping.

\paragraph{Alignment.}
Alignment reconciles differences in schemas and vocabularies so that equivalent concepts mean the same thing at both ends. 
Translators remap structure (fields and nesting), normalize values (units, time zones, encodings), and resolve identities across namespaces.
Mappings are derived from published contracts and shared dictionaries, are versioned and deterministic, and include an explanation and a rule identifier for audit purposes. 
The translated payload is then validated against the target’s schema and invariants. If a mapping is ambiguous or information-losing, the layer requests clarification or escalates to a human, as per policy. 

\paragraph{Verification.}
Verification turns plans and results into checked commitments. The layer evaluates whether a step is permitted and correct given the intended outcome and current state: it enforces pre/postconditions and invariants, applies policy constraints, and validates units, ranges, identities, and time semantics. Outputs must either logically follow from the inputs or be supported by cited evidence; interactive claims may require justification or machine-checkable proofs. When results are probabilistic, the layer attaches calibrated confidence and residual risk and exposes thresholds for escalation or rework. Each check yields a structured verdict (pass/fail, reasons, fix hints) with timestamps and provenance for audit; failures trigger repair suggestions or safe rollback per policy.

\paragraph{Meaning-aware security and guardrails.} Security here protects and governs actions based on what data and operations represent, not just how they are transmitted. Artifacts carry classification and provenance, and policies are expressed in terms of concepts and capabilities. Guardrails enforce intent whitelists, validate arguments against schemas and invariants (units, ranges, preconditions), sandbox untrusted inputs, and check tool outputs before reinjection, mitigating prompt injection and misuse. Every step records lineage and justification, ensuring that decisions are auditable and reproducible. Capability scopes limit what an agent can do, even if language guidance is subverted.

\section{Protocols}
\label{sec:protocol}

\begin{table*}[ht]
\centering
\small
\setlength{\tabcolsep}{4pt}
\renewcommand{\arraystretch}{1.15}
\begin{tabular}{l|ccc|ccc|ccc}
\hline
\multirow{2}{*}{Protocol} 
& \multicolumn{3}{c|}{Communication} 
& \multicolumn{3}{c|}{Syntactic} 
& \multicolumn{3}{c}{Semantic} \\
& Transport & Stream & Security 
& Schema & Lifecycle & Error 
& Clarify & Context & Verify \\
\hline

MCP                & $\checkmark$ & $\checkmark$ & $\triangle$ & $\checkmark$ & $\triangle$ & $\checkmark$ & $\times$ & $\times$ & $\times$ \\
agents.json        & $\checkmark$ & $\times$      & $\checkmark$ & $\checkmark$ & $\times$      & $\triangle$ & $\times$ & $\triangle$ & $\times$ \\
Agent Protocol     & $\checkmark$ & $\times$      & $\triangle$ & $\checkmark$ & $\checkmark$ & $\checkmark$ & $\times$ & $\checkmark$ & $\times$ \\

\hline

ACP-AGNTCY         & $\checkmark$ & $\checkmark$ & $\checkmark$ & $\checkmark$ & $\checkmark$ & $\checkmark$ & $\times$ & $\checkmark$ & $\triangle$ \\
ACP-IBM            & $\checkmark$ & $\checkmark$ & $\checkmark$ & $\checkmark$ & $\checkmark$ & $\triangle$ & $\times$ & $\triangle$ & $\times$ \\
ACP-AgentUnion     & $\checkmark$ & $\times$      & $\checkmark$ & $\checkmark$ & $\triangle$ & $\triangle$ & $\times$ & $\triangle$ & $\triangle$ \\
ACN                & $\checkmark$ & $\triangle$ & $\checkmark$ & $\triangle$ & $\triangle$ & $\triangle$ & $\times$ & $\triangle$ & $\triangle$ \\

\hline

A2A                & $\checkmark$ & $\checkmark$ & $\checkmark$ & $\checkmark$ & $\checkmark$ & $\checkmark$ & $\checkmark$ & $\checkmark$ & $\triangle$ \\
Agora              & $\checkmark$ & $\triangle$ & $\checkmark$ & $\checkmark$ & $\triangle$ & $\triangle$ & $\checkmark$ & $\checkmark$ & $\checkmark$ \\
LOKA               & $\triangle$ & $\triangle$ & $\triangle$ & $\triangle$ & $\triangle$ & $\triangle$ & $\triangle$ & $\triangle$ & $\triangle$ \\

\hline

ANP                & $\checkmark$ & $\triangle$ & $\checkmark$ & $\checkmark$ & $\triangle$ & $\triangle$ & $\times$ & $\checkmark$ & $\checkmark$ \\
Coral              & $\checkmark$ & $\triangle$ & $\checkmark$ & $\checkmark$ & $\checkmark$ & $\triangle$ & $\triangle$ & $\checkmark$ & $\triangle$ \\
AITP               & $\triangle$ & $\triangle$ & $\triangle$ & $\triangle$ & $\triangle$ & $\triangle$ & $\triangle$ & $\triangle$ & $\triangle$ \\

\hline

LMOS               & $\checkmark$ & $\triangle$ & $\checkmark$ & $\checkmark$ & $\triangle$ & $\triangle$ & $\triangle$ & $\checkmark$ & $\checkmark$ \\
CrowdES            & $\times$      & $\times$      & $\times$      & $\times$      & $\times$      & $\times$      & $\times$      & $\times$      & $\times$ \\
SPPs               & $\triangle$ & $\times$      & $\times$      & $\triangle$ & $\triangle$ & $\triangle$ & $\times$ & $\triangle$ & $\checkmark$ \\
PXP                & $\checkmark$ & $\times$      & $\triangle$ & $\checkmark$ & $\checkmark$ & $\checkmark$ & $\checkmark$ & $\checkmark$ & $\checkmark$ \\
WAP                & $\triangle$ & $\triangle$ & $\triangle$ & $\triangle$ & $\triangle$ & $\triangle$ & $\triangle$ & $\triangle$ & $\triangle$ \\

\hline
\end{tabular}
\caption{Comparative analysis of 18 representative agent communication protocols under the proposed three-layer taxonomy. The communication layer covers transport, streaming, and security; the syntactic layer covers schema definition, lifecycle management, and error handling; and the semantic layer covers clarification, context alignment, and verification. $\checkmark$ indicates explicit protocol-level support, $\triangle$ indicates partial or implementation-dependent support, and $\times$ indicates little or no explicit support.}
\label{tab:protocol-coverage}
\end{table*}

To provide a comparative overview of the current landscape, Table~\ref{tab:protocol-coverage} evaluates 18 representative agent communication protocols through the three-layer taxonomy introduced in Section~\ref{sec:taxonomy}. Rather than reproducing protocol specifications feature by feature, the table abstracts each protocol into a common set of dimensions, enabling systems with different design goals to be compared within a single analytical framework.

We assess each protocol along nine dimensions grouped into three layers. The communication layer covers transport, streaming, and security; the syntactic layer covers schema definition, lifecycle management, and error handling; and the semantic layer covers clarification, context alignment, and verification. These dimensions are intended to capture whether a protocol supports reliable exchange, structured interaction, and meaning-level coordination, respectively.

We use a coarse-grained qualitative scoring scheme. A checkmark denotes explicit protocol-level support, a triangle denotes partial, indirect, or implementation-dependent support, and a cross denotes little or no explicit support in the specification. The scores are based primarily on official protocol specifications and publicly stated design goals rather than on individual downstream implementations. Accordingly, Table~\ref{tab:protocol-coverage} should be read as a comparative interpretive summary of protocol design emphasis, not as a benchmark leaderboard.

Three high-level patterns emerge from Table~\ref{tab:protocol-coverage}. First, most protocols are comparatively mature at the communication layer: they usually define a transport substrate and often include at least some support for streaming or security. Second, syntactic support is also relatively strong across the ecosystem, especially for schema specification and structured interaction patterns such as task, session, or capability lifecycles. Third, and most importantly, semantic support remains sparse and uneven. Only a small subset of protocols, such as A2A, Agora, and PXP, provide relatively explicit support for clarification, context alignment, or verification, whereas many others leave these responsibilities to application-level logic.

This imbalance is central to our argument. The existing protocol design has largely succeeded in standardizing how agents connect and how messages are structured, but has been less successful in standardizing how agents resolve ambiguity, maintain shared meaning, and validate interpretations. In other words, many protocols help agents exchange messages correctly without ensuring they understand them in the same way.

This gap has important architectural consequences. When semantic mechanisms are absent from the protocol itself, developers must reintroduce them through prompts, orchestration logic, wrappers, or task-specific adapters. Such workarounds may be practical in the short term, but they also accumulate hidden complexity and reduce interoperability. We return to this issue in Section~\ref{sec:techdebt}, where we analyze it as a form of semantic technical debt.

Table~\ref{tab:protocol-coverage} also provides a roadmap for the remainder of this section. In the following subsections, we examine representative protocols in greater detail, showing how different design philosophies prioritize different parts of the communication, syntactic, and semantic stack.

\subsection{MCP} 
\label{subsec:mcp}
The \textbf{Model Context Protocol (MCP)} is one of the most influential early attempts to standardize communication among LLM-powered agents and external systems. 
Originally proposed by Anthropic, MCP is designed to provide a lightweight yet extensible mechanism for connecting language models to external tools, APIs, and knowledge sources in a structured way. 
Its core philosophy is to make the context around a model, such as tools, data, and memory, easily discoverable and callable through a well-defined interface, thereby enabling agents to interact more intelligently with their environment without needing extensive prompt engineering or custom glue code.

Before diving into technical details, it is important to clarify the roles defined by MCP: 
\textbf{MCP} itself is a protocol specification and standard that defines how agents communicate with external services.
\textbf{MCP Server} is an implementation role that registers available tools, APIs, and functions, exposing them to the ecosystem.
\textbf{MCP Client} is typically an LLM agent that discovers the MCP Server’s capabilities and invokes them via standardized calls.
With these roles in mind, we can better understand how MCP is structured.
Technically, MCP adopts a minimalist architecture composed of three main components: \textbf{(1)} a transport mechanism that connects agents and services; \textbf{(2)} a message contract that specifies the shape, format, and direction of messages; and \textbf{(3)} a capability registration mechanism that allows services to describe the tools and functions they expose. 
By design, MCP focuses on the lower two layers of the agent communication stack: \textbf{communication} and \textbf{syntax}, while intentionally leaving the \textbf{semantic layer} largely unspecified. 
This design choice reflects MCP’s emphasis on interoperability and simplicity, but it also reveals its limitations when it comes to deeper semantic alignment and intent understanding between agents.

\subsubsection{Communication Layer}
At the communication layer, MCP ensures that information can be reliably exchanged between agents and services, and it also takes responsibility for managing the entire connection lifecycle. 
This includes connection establishment; such as handshakes and initialization; capability negotiation, including advertising and retrieving available tools; and orderly teardown when communication ends. 
It primarily supports two transport mechanisms: 
\textbf{(1) Standard Input/Output (stdio):} 
A lightweight communication channel where the model or agent process communicates directly with the MCP service via standard input and output streams. 
This method is especially suited for local deployments or cases where minimal infrastructure is desired.
\textbf{(2) Streamable HTTP:} 
A network-based transport option that uses HTTP to establish streaming communication between the client (an LLM agent) and the server (tool provider).
This enables persistent, low-latency interactions and supports real-time data transfer.

Both methods are designed to ensure that data transmission is consistent, bidirectional, and observable. 
MCP provides basic guarantees on message delivery and ordering, allowing agents to maintain coherent conversational state across multiple requests. 
However, it does not prescribe higher-level features like retries, congestion control, or security policies, leaving such concerns to the deployment environment. 
These aspects, however, can be addressed by integrating MCP with existing solutions, for example, using OAuth to obtain access tokens, attaching API keys or custom headers for authentication, or leveraging HTTP header-based security policies.

\subsubsection{Syntactic Layer}
At the syntactic layer, MCP specifies the structural conventions for how agents and external services construct, exchange, and interpret messages.
It builds upon the widely adopted \textbf{JSON-RPC 2.0} standard, which defines a compact yet expressive schema for remote procedure calls.
Typical message objects contain fields such as \texttt{method}, \texttt{params}, \texttt{id}, and \texttt{result}/\texttt{error}, enabling consistent parsing and interpretation on both sides of the communication.

Beyond defining message structure, MCP also specifies how these messages are exchanged in a broader interaction workflow.
A defining feature of MCP’s design is its strict separation of roles in the interaction workflow. 
The \textbf{MCP server} acts as a capability provider: it publishes a list of available functions, APIs, or contextual data sources that it can expose to the ecosystem. 
The \textbf{MCP client}, usually an LLM-powered agent, plays the role of capability consumer: it can query this registry, invoke specific functions, and process returned outputs. 
Because capabilities are described declaratively and accompanied by schemas, clients can automatically explore what is available without prior hard-coding, verify the validity of their calls, and interpret the results in a uniform manner.
This “directory-like” discovery mechanism is central to MCP’s philosophy: services should describe themselves in sufficient detail so that agents can flexibly integrate new capabilities without manual orchestration.

The clear delineation of roles and responsibilities has two major consequences. 
First, it reduces the ambiguity surrounding how functions are invoked and how responses are structured, which significantly simplifies interoperability across heterogeneous systems. 
Second, the inclusion of standardized error representations, status codes, and schema-based validation enables machine agents to reason about failure conditions and build robust recovery strategies. 
Together, these elements allow MCP to function as a stable syntactic interface layer, ensuring that different components of an agentic ecosystem can interact seamlessly even when built by independent parties.

\subsubsection{Semantic Layer}
While MCP establishes strong foundations for communication and syntax, it does not define or guarantee semantic understanding between agents. 
The protocol focuses primarily on message transport and structure, leaving the interpretation of content, intent, and contextual meaning to the higher-level logic implemented by the agents themselves.
In practice, this means that even if two agents can successfully exchange JSON-RPC messages over a stable channel, there is no built-in mechanism to ensure they interpret the exchanged information in the same way. 
For example, a tool may expose a function named \texttt{book\_flight}, but MCP does not provide any ontology, schema mapping, or intent negotiation mechanism to guarantee that the calling agent fully understands its expected behavior or broader context. 
The MCP server does offer partial support by describing input/output formats, parameter constraints, and validation logic through JSON Schema, but this remains a syntactic-level guarantee rather than a semantic understanding. 
It cannot, for instance, ensure that both sides share the same notion of what a “flight booking” entails or how edge cases should be resolved.
This limitation becomes particularly evident in multi-agent systems, where robust collaboration often requires proactive clarification, semantic alignment, and shared conceptual grounding.

Taken together, MCP can be viewed as a protocol that ensures agents can \textit{“hear”} one another (communication layer) and that they \textit{“speak the same language”} (syntactic layer), but it provides no guarantees that they truly \textit{“understand each other’s intentions”} (semantic layer). 
This gap mirrors the human communication analogy introduced in Figure~\ref{fig:agents-vs-human}: just as successful conversation requires more than audible sound and shared vocabulary, robust multi-agent collaboration demands mechanisms for resolving ambiguity, negotiating meaning, and aligning on shared context — areas where MCP remains minimal. 
Accordingly, within our three-layer evaluation framework, MCP shows strong support at the communication layer through standardized transports such as stdio and streamable HTTP, and at the syntactic layer through JSON-RPC, schema-based capability registration, and structured error handling; however, it remains weak at the semantic layer because it does not define protocol-level mechanisms for clarification, shared context maintenance, or intent verification. 
This overall assessment is summarized in Table~\ref{tab:protocol-coverage}.

\subsection{Agents.json (WildCard AI)} 
\label{subsec:agents-json}

Unlike fuller agent protocols such as MCP or Agent Protocol, \texttt{agents.json} is not intended to standardize the entire lifecycle of agent interaction. 
It does not define a transport channel, session model, or agent-runtime interface. 
Instead, its scope is deliberately narrower: it specifies how external API capabilities can be described in a form that LLM agents can discover, interpret, and compose. 
For this reason, \texttt{agents.json} is best understood not as a full-stack protocol, but as a declarative capability-specification layer. 
It is nevertheless important to compare it with fuller protocols, because it highlights a different design choice in the protocol landscape: rather than standardizing how agents communicate end to end, it standardizes how actionable operations are exposed to agents in an LLM-native way.

This narrower scope addresses a practical gap left by communication-oriented protocols. 
Standards such as MCP mainly define how agents connect to external tools and exchange structured messages, but they leave largely open how those tools should be packaged as reusable, machine-readable capabilities for multi-step use. 
In practice, many APIs already have OpenAPI specifications, yet these are primarily written for human developers and typically assume manual orchestration, procedural reasoning, and implicit contextual knowledge. 
The \textbf{agents.json} specification\footnote{\url{https://github.com/wild-card-ai/agents-json}}, introduced by WildCard AI\footnote{\url{https://wild-card.ai/}}, reworks this layer for LLM agents. 
It defines declarative structures such as \texttt{flows}, \texttt{actions}, and \texttt{links}, allowing APIs to be represented not merely as isolated endpoints, but as executable, composable workflows that an agent can follow directly. 
To operationalize these manifests, WildCard also provides a component called the \textbf{Bridge}, which the project documentation describes as a Python package for loading, parsing, and running \texttt{agents.json} workflows, with quickstart examples covering API key, bearer-token, and OAuth-based authentication.
In this sense, MCP and Agent Protocol can be viewed as fuller protocols for communication and lifecycle management, whereas \texttt{agents.json} contributes a more specialized but complementary layer: describing what external capabilities are available and how they can be composed for agent use.

\subsubsection{Communication Layer}
At the communication layer, \texttt{agents.json} takes a minimalist stance. 
It does not define any new transport protocol, persistent connection model, or session-handling mechanism. 
Instead, execution is delegated to the underlying infrastructure that already supports APIs today—most commonly HTTP, HTTPS, or service-specific SDKs. 
Each invocation of a flow or action is \textbf{stateless}: all required parameters must be provided, 
and no shared session context is assumed between consecutive calls. 
This aligns with the paradigm of serverless and pub/sub architectures, where actions are atomic and self-contained.

In practice, execution is supported through \textit{Bridge}, which the project documentation describes as tooling for loading, parsing, and running \texttt{agents.json} workflows over existing API integrations.
Because this layer reuses the mature security and orchestration mechanisms of existing APIs (API keys, OAuth tokens, or HTTPS), \texttt{agents.json} can immediately integrate into production systems without introducing additional transport complexity. 
Compared with MCP’s bi-directional streaming design (via stdio or HTTP streaming), 
\texttt{agents.json} focuses on a one-shot, declarative invocation model that favors modularity and composability over persistent state.

\subsubsection{Syntactic Layer}
The syntactic layer is where \texttt{agents.json} contributes its main innovation. 
It defines a structured and strongly typed representation of how an agent can execute external APIs, 
building upon the OpenAPI standard but extending it beyond individual endpoints to describe complete workflows. 
Central to this layer are three core constructs: \texttt{flows}, \texttt{actions}, and \texttt{links}. 
A \texttt{flow} represents a goal-oriented task that can be decomposed into multiple steps. 
Each step, or \texttt{action}, corresponds to a single API call, typically referencing an OpenAPI \texttt{operationId} that specifies how that call is made. 
While actions describe \emph{what to invoke}, links describe \emph{how information moves} across them. 
A \texttt{link} connects the output of one action to the input of another or maps high-level flow parameters to specific arguments within an API call. 
Through these constructs, \texttt{agents.json} captures both the structural logic and the data dependencies of a task in a purely declarative way.

Each \texttt{flow} specifies a goal-oriented multi-step task, consisting of one or more \texttt{actions} connected by \texttt{links}. 
For example, a Google Sheets integration may define a flow that (1) creates a sheet, (2) appends a row, and (3) generates a share link. 
A corresponding \texttt{link} ensures that the output field \texttt{response.spreadsheetId} from the first action is passed as the input parameter 
\texttt{parameters.spreadsheetId} to the next. 
Links can represent both \emph{sequential} dependencies—passing outputs between actions—and \emph{parallel} mappings, 
such as propagating multiple input parameters from the flow-level definition into a single API call. 
This declarative linking mechanism allows \texttt{agents.json} to represent workflow-like dependency structures without imperative orchestration code.

Because each action ultimately points back to a formal OpenAPI definition, 
the overall structure inherits OpenAPI’s precision while adding a higher level of compositional logic. 
All flows and links are validated through machine-readable JSON Schema rules, 
ensuring that references, types, and field paths remain consistent. 
In this sense, \texttt{agents.json} functions as a meta-layer on top of OpenAPI—%
transforming static API documentation into executable agent manifests that LLMs can interpret and orchestrate directly.

\subsubsection{Semantic Layer}
At the semantic layer, \texttt{agents.json} remains intentionally lightweight. 
It does not prescribe any ontology, formal intent representation, or shared vocabulary. 
The meaning of each capability is inferred primarily from natural-language descriptions embedded within the schema and the semantic cues of action names. 
Thus, semantic alignment—ensuring that two agents interpret a capability in the same way—remains an open challenge. 
Nevertheless, the inclusion of \texttt{flows} and \texttt{links} introduces a weak form of semantic structure: 
they define goal-directed decompositions of tasks and explicit data dependencies, 
providing a bridge between symbolic planning and API invocation. 
Future extensions may enrich this layer with concept alignment mechanisms or shared intent ontologies, 
allowing \texttt{agents.json} manifests to serve as semantically grounded building blocks for interoperable agent ecosystems.

Conceptually, MCP provides the \textit{“how to talk”} framework for agent communication, 
whereas \texttt{agents.json} defines \textit{“what can be done”} within those conversations. 
Their combination—using \texttt{agents.json} manifests as callable capabilities within MCP-style transport layers—%
offers a promising foundation for an extensible, safe, and interoperable “Internet of Agents.”

\subsection{Agent Protocol (AGI, Inc.)} 
While the Model Context Protocol (MCP) and related standards focus on how language models call tools or attach context, 
their primary role is still to wrap \emph{a single agent} or model behind a structured interface. 
They say little about how to expose the full lifecycle of an autonomous agent itself: 
how to start a task, advance it step by step, inspect its internal progress, or plug that agent into generic devtools and benchmarks. 
In practice, each agent framework today typically defines its own bespoke HTTP or WebSocket API, 
making it difficult to swap agents, compare them fairly, or build reusable dashboards and orchestration layers on top.

The \textbf{Agent Protocol}\footnote{\url{https://github.com/agi-inc/agent-protocol}}, maintained by AGI, Inc.\footnote{\url{https://www.theagi.company/}},  is designed precisely to address this gap. 
Rather than being tied to any specific framework or runtime, it specifies a small, tech-stack–agnostic REST API (defined in an OpenAPI specification) that every compatible agent is expected to implement.\footnote{\url{https://agentprotocol.ai/}}\footnote{\url{https://agent-protocol-docs.vercel.app/}} 
At its core, the protocol standardizes a handful of concepts: 
\emph{tasks} that capture high-level objectives, 
\emph{steps} that record incremental progress toward those objectives, 
and auxiliary resources such as artifacts. 
By unifying these lifecycle operations, Agent Protocol makes it possible for external ``control consoles'' 
to start, monitor, and analyze heterogeneous agents in a uniform way, 
opening the door to common benchmarking suites, observability tooling, and multi-agent orchestrators.

\subsubsection{Communication Layer}
At the communication layer, Agent Protocol adopts a deliberately conservative design. 
It does not introduce a new transport substrate or streaming wire format; instead, it assumes ordinary HTTPS JSON APIs as the underlying channel. 
Every interaction with an agent is mediated through a small set of REST endpoints under the \texttt{/ap/v1/agent} namespace, 
typically including:

\begin{itemize}[leftmargin=1.3em]
    \item \texttt{POST /ap/v1/agent/tasks} to create a new task (e.g., ``refactor this codebase'' or ``research this topic''),
    \item \texttt{POST /ap/v1/agent/tasks/\{task\_id\}/steps} to advance that task by one reasoning or action step,
    \item auxiliary endpoints for listing tasks and steps, retrieving task details, and uploading, listing, or downloading artifacts.\citep{alengineerfoundation_agent_2025}
\end{itemize}

Each HTTP request is stateless at the transport level, but the protocol exposes task- and step-level identifiers that allow clients to revisit and monitor ongoing work across requests. In production deployments, the public documentation recommends storing task and step state in an external database rather than relying on in-memory storage.
Reliability and security are delegated to the underlying web infrastructure (TLS, authentication headers, rate limiting), 
which allows Agent Protocol to integrate into existing production environments without special transport support. 
Compared with MCP's emphasis on low-latency bidirectional streaming, 
Agent Protocol is closer to a workflow engine: 
it favors durable, inspectable task state over fine-grained token streaming.

\subsubsection{Syntactic Layer}
The syntactic layer is where Agent Protocol does most of its work. 
It defines not only individual endpoints, but also a shared vocabulary of JSON objects and fields 
that describe an agent’s behavior in a framework-independent way. 
The core schemas include:

\begin{itemize}[leftmargin=1.3em]
    \item a \textbf{Task} object, which captures a high-level objective together with identifiers, input fields, and associated artifacts, and may also expose lifecycle metadata depending on the implementation;
    \item a \textbf{Step} object, which records one unit of progress on a task, typically including identifiers, status, output fields, and any artifacts produced at that step;
    \item optional \textbf{Artifact} objects, which reference files or auxiliary outputs produced during execution.
\end{itemize}

Responses are expected to conform to the protocol’s JSON response models and standard HTTP status conventions, allowing generic clients to inspect and visualize compatible agents in a uniform way.
This means that a generic client can inspect and visualize the execution of any Agent-Protocol–compatible agent 
without knowing which internal framework (Auto-GPT, custom planner, etc.) is driving it.\citep{alengineerfoundation_agent_2025} 
From the perspective of our taxonomy, 
the protocol therefore offers a strong syntactic contract: 
it fixes how tasks, steps, and artifacts must be represented on the wire, 
while intentionally remaining agnostic to how individual agents implement planning, memory, or tool use internally.

\subsubsection{Semantic Layer}
At the semantic layer, Agent Protocol remains intentionally lightweight. 
The protocol does not prescribe a formal ontology of goals, 
nor does it attempt to define the meaning of specific task types (e.g., ``code\_review'' vs.\ ``data\_analysis''). 
The semantics of a task are largely encoded in natural-language fields (such as the task description) 
and in whatever domain-specific metadata an agent chooses to attach. 
Consequently, two agents may both implement the protocol correctly at the communication and syntactic levels, 
yet still interpret similar task descriptions in quite different ways.

That said, the Task/Step abstraction introduces a weak but useful layer of semantic structure. 
A task represents a coherent objective; 
steps represent an ordered reasoning or action trace toward that objective; 
artifacts represent durable byproducts. 
This encourages agents to expose a transparent execution trail rather than a single opaque response, 
and enables external tools to reason about progress (e.g., ``how many steps were needed'', ``where did failures occur'') in a protocol-agnostic manner. 
In long-term deployments, one can imagine higher-level conventions emerging on top of these building blocks 
(e.g., agreed-upon task types, standard metadata fields for safety or compliance), 
but these are outside the current specification and left to ecosystem practice rather than enforced by the protocol itself.

Conceptually, where MCP primarily standardizes \textit{how models and tools talk} 
and \texttt{agents.json} specifies \textit{what operations those tools can perform}, 
Agent Protocol focuses on \textit{how full agents are driven and observed over time}. 
It turns autonomous agents into first-class, inspectable services with a common lifecycle interface, 
which in turn makes it easier to benchmark, monitor, and orchestrate heterogeneous agent implementations within a larger ``Internet of Agents.'' 
\subsection{ACP-AGNTCY} 
\label{subsec:acp-agntcy}
The Agent Connect Protocol (ACP) is part of Cisco Outshift's AGNTCY ecosystem, and is now hosted under the Linux Foundation.
The protocol distinguishes itself through a strictly event-driven, schema-aware architecture optimized for high-performance, real-time multi-agent collaboration.
Unlike standard request-response protocols, ACP-AGNTCY treats interactions as continuous event streams, allowing agents to emit incremental outputs, invoke tools dynamically, and perform nested agent-to-agent calls without breaking connection state.
Its syntactic layer enforces a unified event taxonomy encompassing lifecycle, message, tool call, and state events.
Semantically, its uniqueness lies in the robust ``Threads'' mechanism for conversational state preservation across sessions and agents, paired with the Open Agent Schema Framework (OASF) for machine-readable capability discovery and routing.
This makes it well-suited for secure, large-scale deployments where temporal context and live streaming are critical.

\subsubsection{Communication Layer}
ACP-AGNTCY uses a client-server communication architecture with an emphasis on asynchronous, event-driven message exchange.
An agent is deployed behind a standard web endpoint, and clients or other agents connect to it using HTTP/HTTPS.
The protocol supports real-time streaming of results, where the agent returns a sequence of events instead of a single response payload.
This is commonly implemented via Server-Sent Events (SSE), which the ACP clients utilize to subscribe to an agent's output stream.
When a client invokes an agent's Run operation, the agent processes the request and streams back various events (e.g., start-of-task, partial results, tool invocations, final result) in order.
This communication layer design accommodates long or multi-step tasks by providing incremental updates.
It also facilitates multi-agent collaboration since an agent can act as a client to call another agent's API.
The ACP's infrastructure supports maintaining context across these calls.
For example, agents can pass a Thread Identifier along with requests so that context is shared and the conversation remains continuous.
While each interaction still happens via a central endpoint, the protocol is scalable and flexible, as agents can be discovered via AGNTCY's registry services and then addressed by name over the network.
In essence, the communication layer allows any compliant agent to be invoked as a web service with streaming results, and it enables agents to talk to each other by making nested calls, all through a uniform event-driven interface.

\subsubsection{Syntactic Layer}
The syntactic layer of ACP-AGNTCY is defined around an event stream model.
Rather than single-request/single-response messages, interactions are broken down into a standardized sequence of typed events, each following a common JSON structure.
Every event emitted by an agent includes an event type field and an associated payload, adhering to a unified scheme for all agents.
Important event types include lifecycle events to denote high-level stages of execution, message events for primary content, tool call events which indicate the agent is invoking an external tool or another agent, and state events for changes in internal state or context sharing.
All such events conform to a basic format and are rigorously type-checked, ensuring that an event has the expected fields and structure across any agent.
The payloads of events are often JSON objects or encoded text, and can also reference binary data to support multimodal content like images or audio.
By enforcing a consistent event syntax, ACP-AGNTCY allows clients to subscribe and react to streams in a uniform way.
This layered syntax abstraction makes it easier to build tools on top of the protocol and to combine capabilities.

\subsubsection{Semantic Layer}
The semantic layer of ACP-AGNTCY focuses on maintaining context and defining the meaning of various interactions so that multiple agents can coordinate effectively.
A key feature is its Threads Mechanism for context management.
Semantically, a Thread in ACP-AGNTCY represents a conversational or task context;
it links a series of messages or events across one or more agents into a coherent session.
Agents can create new Threads for independent tasks, copy or fork Threads when delegating a subtask to another agent, and even query or search Thread histories.
This means that the protocol inherently supports context continuity.
The content of messages is interpreted in light of the Thread's prior events, enabling multi-turn dialogues or iterative workflows where each agent message builds on the last.
Moreover, the AGNTCY ecosystem pairs ACP with the OASF for agent descriptors, which provides a semantic description of an agent's capabilities, inputs, and outputs in a standardized format.
This allows an agent to advertise what it can do in a machine-readable way, and it complements the ACP protocol by enabling automated discovery and coordination.
In summary, the semantic layer of ACP-AGNTCY ensures that beyond low-level message exchange, agents share an understanding of the context of the conversation and meaning of events, and it leverages a common agent ontology to align their higher-level interactions.

\subsection{ACP-IBM} 
\label{subsec:acp-ibm}
The IBM Agent Communication Protocol (ACP-IBM) features a minimalist, web-native approach to multi-agent interoperability.
In stark contrast to heavy, stateful protocols, ACP-IBM is designed for frictionless integration, treating every agent---whether an LLM, a simple tool wrapper, or a microservice---as an easily accessible REST-style web service.
A defining feature of ACP-IBM is its elegant message schema centered on "roles" and multi-modal "Parts", which allows agents to seamlessly exchange text, images, audio, or artifacts within a unified envelope without requiring complex payload parsing.
Furthermore, it natively supports a "router agent" topology to mediate complex workflows and task distribution.
By intentionally avoiding a rigid semantic ontology and instead relying on Agent Manifests or capability advertisement, ACP-IBM prioritizes rapid ecosystem adoption, composability, and straightforward asynchronous streaming via Server-Sent Events.

\subsubsection{Communication Layer}
ACP-IBM employs a centralized client-server communication model using standard web protocols.
Agents expose a RESTful HTTP API for messaging, enabling interoperability without custom transport layers.
The protocol supports asynchronous, streaming interactions by default while also allowing synchronous calls for low-latency needs.
Each agent instance is identified by an Agent ID and can be discovered via multiple methods, including direct invocation, registry-based lookup, or offline metadata embedded in agent packages.
In simple cases, a client connects directly to an agent's HTTP endpoint to exchange messages.
For complex workflows involving multiple agents, IBM's ACP introduces an optional router agent that can coordinate tasks, which mediates messages among agents to fulfill a multi-agent request.
The communication layer includes built-in support for long-running tasks, while an agent can return intermediate results via streaming through asynchronous calls.
Agents can also authenticate callers and maintain context across messages, preserving state for multi-turn interactions.
Overall, ACP-IBM's communication layer prioritizes framework-agnostic connectivity using simple web standards, with mechanisms for discovery, routing, and streaming that make agent integration seamless.

\subsubsection{Syntactic Layer}
The ACP-IBM protocol defines a clear JSON-based message format for structuring agent interactions.
Every message contains a defined role and a list of Parts that hold the content payload.
Each Part includes a content field, optionally with a URL for large or binary content, enabling multi-modal communication without altering the message structure.
ACP-IBM's syntax also allows metadata within messages.
For example, Parts can carry a metadata object for semantic annotations like citations or additional context.
All messages and events conform to the same base schema, which simplifies parsing and validation.
By using a unified JSON structure with content-type tagging, the syntactic layer ensures that different agents and tools can understand the format of data being exchanged and handle it appropriately.
This consistency reduces integration efforts and errors, as any ACP-compliant agent will produce and consume messages in this expected format.

\subsubsection{Semantic Layer}
ACP-IBM does not enforce a particular ontology for message content; agents often communicate using natural language or domain-specific JSON within the message Parts.
However, the protocol does provide a semantic framework for agent capabilities.
Each agent comes with an Agent Manifest advertising its identity, description, and supported input/output types.
This manifest serves as a semantic contract so that others understand what the agent can do, while the protocol leaves the interpretation of the message content itself to the agents' logic or the natural language embedded in those messages.

\subsection{ACP-AgentUnion} 
ACP-AgentUnion proposes a radically decentralized, Internet-scale vision for multi-agent communication, conceptualizing agents as globally addressable entities akin to standard websites.
Its most unique architectural choice is the Agent Identifier (AID) system, which assigns agents a resolvable domain name hosted by an Access Point (AP) node that handles routing, load balancing, and authentication.
This creates a peer-to-peer federated mesh over standard HTTPS, bypassing centralized bottlenecks.
Beyond its robust syntactic layer for JSON-based request-response schemes, ACP-AgentUnion stands out in its semantic layer by integrating built-in standards for authorization and economic transactions.
Agents can automatically negotiate paid services, enforce usage limits, and deduct credits for API usage.
Combined with AP-exposed capability metadata indexed by agent search engines, this protocol forms a trustable, scalable foundation for cross-organizational and commercial agent collaboration.

\subsubsection{Communication Layer}
ACP-AgentUnion’s communication layer is designed as a decentralized network of agents.
Each agent is given a globally unique identifier (AID) in the form of a domain name, which serves as its address in the ``agent network”.
In practice, an agent registers its name with an Access Point (AP) server in the network; the AP provisions a second-level domain (e.g. \texttt{<agent-name>.<ap-domain>}) as the agent’s AID and hosts the agent’s connectivity.
Communication between agents is then achieved via standard HTTPS requests routed through these APs.
When Agent A wants to communicate with Agent B, it will send a message to its own AP, which uses the DNS-resolvable AID of Agent B to locate B’s AP, and then forwards the message to Agent B.
This effectively creates a peer-to-peer mesh of agent servers: there is no single centralized server for all agents, but rather a federation of AP nodes that cooperate to deliver messages.
The communication layer handles important network functions such as authentication and encryption (leveraging HTTPS/TLS), message routing based on the agent’s domain name, and load balancing.
In fact, AP servers are designed to scale horizontally – they can manage sessions and queue messages for many agents concurrently, and built-in load balancing ensures that high volumes of traffic can be handled reliably.
Agents remain reachable as long as their AP is online and their AID is known; discovery is inherent in the system via DNS lookups and agent directories.
This architecture achieves decentralization while still providing structured connectivity: any agent can directly address any other agent by its AID.
Thus, ACP-AgentUnion’s communication layer establishes an “Internet of Agents,” where heterogeneous agents communicate over the web’s infrastructure with global addressing and federation.

\subsubsection{Syntactic Layer}
The syntactic layer of ACP-AgentUnion defines a uniform format for messages and agent metadata to ensure inter-operability across different platforms and programming languages.
All agent-to-agent communication occurs via standard web requests (e.g., RESTful APIs over HTTPS), with data typically serialized as JSON.
The protocol specification outlines how an agent call should be constructed: for instance, an HTTP request might include the target agent’s AID in the URL and a JSON body containing the action or query and any parameters or content required.
Similarly, responses are sent back as JSON with a defined structure for results or errors.
ACP-AgentUnion also specifies the structure of agent metadata records that APs expose, which is essentially a schema for an agent’s profile (including its name, description, and possibly declared capabilities or interface) that can be retrieved or indexed by agent search services.
This means when an agent registers with an AP, it provides information in a standard format which others can later fetch via a directory API or a well-known URL at the AP to learn about that agent.
The ``data specification" part of ACP-AgentUnion covers these formats, ensuring that all agents describe their input/output types or endpoints in a consistent JSON schema.
Another syntactic aspect is session and context handling: messages may carry a session ID or thread ID if part of a longer conversation, and the protocol defines where and how such context identifiers appear in the message structure to be consistently interpreted by APs and receiving agents.
By using JSON and web-standard encodings, ACP-AgentUnion’s syntax is both human-readable and machine-parseable, leveraging existing tools.
In essence, the syntactic layer acts as the grammatical rules of the agent communication “language,” covering both the message exchange format and the public description format for agents in the network.

\subsubsection{Semantic Layer}
ACP-AgentUnion provides a rich semantic layer aimed at enabling trust, discoverability, and cooperation among agents on a global scale.
Central to this is the concept of a unified agent identity and its associated semantics: an agent’s AID (domain name) isn’t just a routing address, but also an identity that can be authenticated and attributed.
The protocol requires that an agent’s identity be confirmed via its Access Point (similar to domain ownership), which gives communicating agents a level of trust that messages are really from the claimed source.
On top of identity, ACP-AgentUnion standardizes how agents describe their capabilities and services.
Each agent’s metadata (shared through the AP) includes a description of what the agent can do, allowing other agents (or a discovery engine) to semantically understand who might fulfill a given task.
The protocol is designed to be search-engine friendly: agent directories can be crawled to build a semantic index of agents by capability.
Beyond service description, ACP-AgentUnion defines norms for authorization and negotiation between agents.
It includes an authorization framework where agents can require credentials or payment tokens to perform certain actions, and a transaction protocol for agents that offer paid services or rate-limited APIs.
These rules ensure that when an agent requests something of another, both sides share an understanding of the terms: e.g., whether the service is free, or what ``credits” to deduct for usage, all governed by a standard workflow in the protocol.
Additionally, AgentUnion’s semantic layer encompasses guidelines for agent behavior and security (for instance, how an agent should handle unrecognized requests, or how to fail gracefully), which are part of the protocol’s aim for reliability and ethical AI interaction.
In summary, ACP-AgentUnion doesn’t just allow agents to exchange bits; it creates a shared understanding of who each agent is (identity), what they can do (capabilities), and how they should cooperate (through agreed authorization and interaction patterns).
This common ground is what enables a diverse, decentralized set of agents to find each other and work together seamlessly under the ACP-AgentUnion standard.

\subsection{ACN} 
The Peer‑to‑peer Agent Communication Network (ACN) diverges from client-server and federated models by offering a fully permissionless, decentralized overlay architecture.
At its core, ACN is structured around a Distributed Has Table (DHT) that maps cryptographically derived agent identifiers directly to contact peers, enabling agents to locate and message each other without any central registry or intermediary.
Its defining strength lies in its profound focus on autonomy and network resilience: it natively supports diverse connection modalities---including direct, relayed (for NAT traversal), and mailbox modes (for offline asynchronous message storage)---to accommodate varied hardware constraints and fluctuating network conditions.
Furthermore, ACN strictly enforces end-to-end encryption and cryptographic identity verification, ensuring that trust and confidentiality are maintained purely at the edge.
This makes ACN the premier choice for highly distributed, privacy-centric, and censorship-resistant multi-agent systems.

\subsubsection{Communication Layer}
At its communication layer, ACN uses a peer-to-peer network of “peers” maintaining a DHT overlay that maps agent identifiers (derived from public keys) to their contact peers.
Each peer functions as both a routing node in the overlay and a proxy for one or more agents.
When an agent wants to send a message, it forwards it to its contact peer, which performs a DHT lookup to locate the destination agent’s contact peer and then establishes a channel (typically over TCP or other transport) to deliver the message.
The system supports different connection modes: a direct connection (agent and peer collocated), a relayed connection (agent behind NAT uses another peer as relay), and a mailbox mode (agent offline, relies on peer for message storage).
Communication is end-to-end encrypted (using public‐key cryptography) and peer identities are verifiable, ensuring confidentiality, authenticity and integrity.
The p2p overlay enables decentralized agent discovery and messaging simply by knowing the target agent’s address (agent id).

\subsubsection{Syntactic Layer}
On the syntactic layer, ACN defines the structure of agent-to-peer and peer-to‐peer protocols and a minimal message envelope.
Each message consists of metadata (originator id, destination id, payload encoding info) plus the payload proper.
The metadata identifies sender and receiver by their agent IDs (hashes of public keys) and indicates how to interpret the payload.
Agents register themselves by signing their contact peer’s ID and publishing the agent-peer association in the DHT.
The lookup and routing protocols are precisely defined: contact peer lookup, peer‐peer routing via the DHT, peer forwarding of messages to the agent.
Various connection types introduce slightly different syntactic flows (e.g., mailbox mode introduces storage/retrieval semantics).
Although the protocol does not prescribe a rich multi-modal payload schema (text, image, etc.), it provides a generic envelope so that arbitrary payload content can be transported.
The key syntactic guarantee is that any peer or agent can parse the header (metadata) and route the message correctly; semantics of the payload are left to agent implementation.

\subsubsection{Semantic Layer}
The semantic layer in ACN focuses on agent identity, autonomy, and trust rather than on formal ontologies of content.
Agents are identified by cryptographic public keys; the address space is derived from those keys, giving each agent a unique and verifiable identity.
That identity semantics allow a sending agent to know who it is connecting with and to authenticate the peer infrastructure; the protocol ensures that agent–peer associations are signed and stored in the DHT, giving verifiable proof of binding.
The peer‐to‐peer overlay semantics also include resilience: the structured DHT (based on S/Kademlia) is chosen for its resilience to malicious behavior and churn, which lends semantic guarantees of availability, correctness of routing, and fault tolerance.
However, ACN does not define high-level semantic layers of conversation (e.g., speech-act types, ontology for tasks).
That is left to the agent frameworks built on top of ACN.
Hence the semantic layer here is about identity, trust, addressing, and guaranteed message delivery semantics, rather than the meaning of the message payload itself.
\newcommand{\hbw}[1]{{\color{green}{[HBW: #1]}}}
\subsection{Agent-to-Agent} 
\label{subsec:a2a}
The Agent-to-Agent (A2A) \cite{google_a2a_2025} Protocol was introduced by Google as a standard for direct communication between autonomous agents across organizations and systems. Its main goal is to enable agents—each potentially built with different models, frameworks, or APIs—to communicate, collaborate, and delegate tasks securely without human intervention or ad-hoc integration code. In short, A2A defines a universal "language" for agents to talk to each other, just as HTTP does for computers.
A2A extends the idea behind the MCP, which mainly connects an agent to tools or APIs. While MCP focuses on the agent-to-tool relationship, A2A generalizes it to agent-to-agent collaboration. It defines how one agent can discover another, negotiate permissions, exchange structured messages, and manage long-running tasks through a shared set of conventions. Conceptually, A2A marks a shift from “a single agent operating within its sandbox” toward “a network of cooperating agents” that form the backbone of an Internet of Agents.
The A2A specification adopts a three-layer architecture---communication, syntactic, and semantic---that maps directly to the taxonomy proposed in this survey. The communication layer ensures connectivity and reliability (\textit{can we talk?}), the syntactic layer standardizes structure and workflow (\textit{are we speaking the same language?}), and the semantic layer enables shared understanding and intent alignment (\textit{do we understand what we mean?}).

\begin{figure}[ht]
\centering
\includegraphics[width=0.9\linewidth]{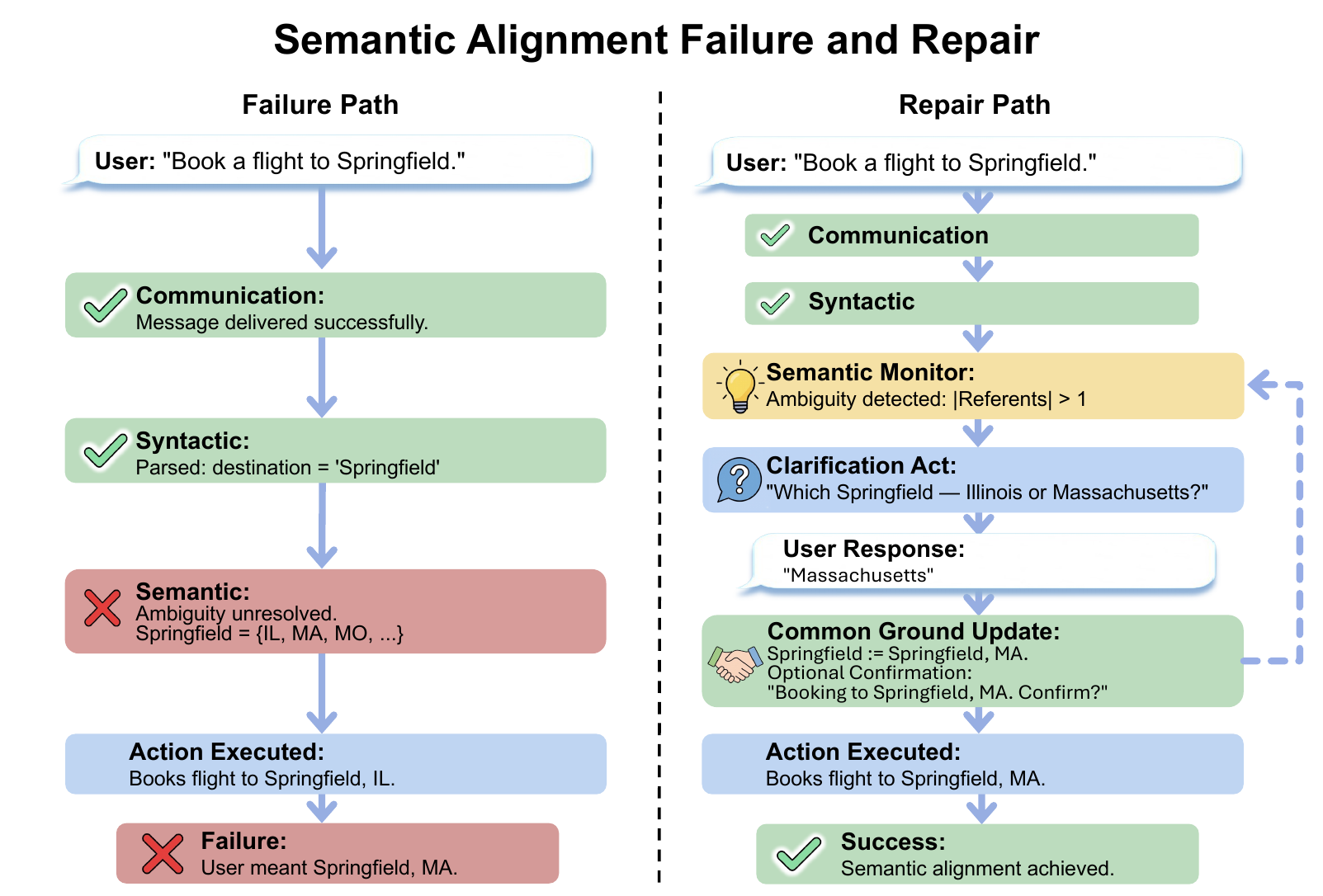}
\caption{Illustrative semantic alignment failure and repair in A2A. After successful communication and syntactic parsing, the request remains semantically ambiguous because “Springfield” may refer to multiple destinations. The repair path detects the ambiguity, triggers a clarification exchange, updates shared context, and executes the task only after intent is aligned.}
\label{fig:a2a}
\end{figure}

As illustrated in Figure~\ref{fig:a2a}, A2A supports semantic repair when a request is communicatively and syntactically valid but semantically ambiguous. Clarification and shared-context updates enable the interacting agents to align intent before task execution.

\subsubsection{Communication Layer}
At the communication layer, A2A specifies how agents discover each other, establish secure connections, and maintain reliable data exchange. Each agent exposes an Agent Card, a small JSON document hosted at \textit{agent-card.json}, containing metadata such as its name, description, capabilities, API endpoints, and required authentication methods. This design parallels web service discovery and allows any agent to learn how to contact another agent simply by retrieving its card.
Once discovery is complete, agents communicate through HTTP/HTTPS as the baseline transport protocol, with optional Server-Sent Events (SSE) for low-latency, streaming updates during long tasks. Each A2A interaction follows a defined lifecycle: discovery $\rightarrow$ authentication $\rightarrow$ message exchange $\rightarrow$progress reporting $\rightarrow$ teardown. The specification recommends the use of OAuth 2.0 bearer tokens or mutual TLS certificates for identity verification. These security measures ensure that both sides know who they are talking to and that information cannot be intercepted or modified.
A2A's communication layer also supports long-running and bidirectional sessions, enabling agents to collaborate beyond one-shot requests. Through SSE or persistent connections, an agent can continuously stream status updates, partial results, or event notifications to its counterpart. This interactive model is particularly important for complex, multi-stage workflows (e.g., when a research agent delegates subtasks to a data-analysis agent). In essence, this layer fulfills the role of "ensuring agents can hear each other clearly and safely," forming the physical and infrastructural backbone for higher-level coordination.
\subsubsection{Syntactic Layer}
The syntactic layer defines how agents organize and interpret messages in A2A. It standardizes the message format, the task structure, and the workflow conventions so that different systems can communicate consistently. Each message follows a uniform schema containing fields such as \texttt{method}, \texttt{params}, \texttt{id}, \texttt{status}, and optional attachments known as artifacts, which can include text, structured data, or files. A2A provides three core interaction methods: \texttt{message/send} for single exchanges, \texttt{message/stream} for continuous updates during long sessions, and \texttt{tasks/get}, \texttt{tasks/list}, or \texttt{tasks/cancel} for managing task lifecycles. Every task carries a unique identifier and explicit states such as pending, working, completed, failed, or cancelled, allowing agents to coordinate long-running processes in a transparent and recoverable way. The syntactic layer also specifies uniform error codes and validation schemas, ensuring that failures and retries are machine-actionable. Although designed primarily for HTTP and JSON, the schema itself is transport independent, meaning that it can be implemented over gRPC, WebSockets, or other channels without loss of meaning. This layer guarantees that agents not only exchange data but also maintain structural consistency, ensuring they speak the same language in every interaction.
\subsubsection{Semantic Layer}
The semantic layer addresses how agents achieve mutual understanding in A2A beyond message syntax. While the protocol does not enforce a universal ontology or fixed vocabulary, it encourages agents to describe their capabilities using both natural language and structured schemas. Each capability listed in an Agent Card includes clear input and output definitions, allowing another agent to interpret what an action means and how it should be invoked. During interaction, if a request is ambiguous or incomplete, the receiving agent can issue a clarification query to confirm intent before execution. This clarification–confirmation process introduces a minimal but effective mechanism for aligning meaning. In more advanced implementations, A2A allows the use of semantic adapters or intent mappers that translate between heterogeneous schemas, enabling agents developed by different organizations to interoperate smoothly. For example, when two agents describe similar actions with different parameter names, these adapters can automatically reconcile the differences. Although A2A's semantic layer remains lightweight, it provides an essential bridge between structured syntax and shared understanding. It transforms communication from simple data transfer into a process of cooperative reasoning, ensuring that agents act not only on the same data but also on the same intent.

\subsection{Agora} 
\label{subsec:agora}
The Agora Protocol \cite{samuele_marro_scalable_2024} defines a meta-protocol that integrates natural-language negotiation with reusable structured routines. It was proposed to address the cost and ambiguity of large-scale LLM-to-LLM communication by allowing agents to gradually converge from unstructured natural-language dialogue to fully specified structured exchanges. Each interaction in Agora is governed by a minimal message contract and, when applicable, a \emph{Protocol Document (PD)}—a text-based specification that formalizes the syntax and semantics of a particular task and is referenced by a content hash. Agents can negotiate and store these PDs, and from them generate deterministic routines to handle repeated queries without additional LLM inference. Through this design, Agora reduces the marginal cost of recurrent communication while maintaining flexibility for novel or one-off tasks. Similar to the structure used for other protocols, the following subsections describe Agora’s communication, syntactic, and semantic layers in detail.

\subsubsection{Communication Layer}

At the communication layer, Agora specifies how agents establish connectivity, exchange messages, and maintain persistent sessions. The protocol relies on standard web transport—typically HTTPS with optional streaming—and defines a lightweight envelope that accompanies every request. Each message contains a \texttt{protocolHash}, a list of \texttt{protocolSources} for retrieving the referenced PD, and a \texttt{body} that carries either structured data or free-form natural language. When an agent receives a request, it checks its local PD cache using the provided hash. If the corresponding PD and routine are available, it executes the routine directly without invoking an LLM. If no match is found, the agent interprets the request with its LLM and, if needed, fetches and stores the PD from the supplied sources for future reuse. The protocol supports ordinary HTTP request lifecycles including opening, processing, and returning success or failure, and it delegates message ordering, retries, and congestion control to the transport layer. Discovery and versioning are handled through a standardized \texttt{/.wellknown} endpoint that exposes supported \texttt{hash→source} mappings, enabling new agents to align quickly with existing protocols. Security is inherited from the transport configuration (TLS, authentication headers, and rotating access tokens). Overall, the communication layer ensures that agents are always able to reach one another—falling back to natural language when necessary—and that once a protocol is shared, subsequent exchanges can proceed deterministically and efficiently.

\subsubsection{Syntactic Layer}

At the syntactic layer, Agora formalizes the message schema and its relation to the referenced PD. Each message follows a uniform structure consisting of a request identifier, timestamps, status indicators, and the \texttt{body} field. When a PD is referenced, the body must conform to the PD-defined input schema, and responses must include outputs that validate against the corresponding output schema. PDs themselves are plain-text specifications describing field names, data types, valid ranges, enumerations, and optional pre- and post-conditions. Each PD is immutable once hashed and may import other PDs, allowing agents to compose complex tasks from simpler sub-protocols. When repeated exchanges for the same PD become frequent, an agent uses its LLM once to synthesize a routine implementing the PD, registers the mapping between the hash and the handler, and then serves future requests through this routine. If a validation error occurs or the routine fails, the responding agent returns a structured error object that includes retry advice and diagnostic context. The caller can then choose to correct the payload and retry, or to initiate a new negotiation by omitting the hash and returning to natural-language dialogue. Through these mechanisms, the syntactic layer provides a stable schema for structured communication while allowing adaptive repair and regeneration when schema drift occurs.

\subsubsection{Semantic Layer}

The semantic layer governs how meaning and intent are aligned and maintained across repeated interactions. During initial negotiation, agents use natural language to propose and clarify field semantics, units, and expected outcomes until a consistent interpretation is reached. This shared understanding is then captured within the PD, including an \emph{intent descriptor} that specifies the purpose of the task, its success conditions, and the guarantees associated with execution. As communication continues, agents monitor usage patterns and automatically generate or retire routines based on frequency and cost thresholds, ensuring that frequently used semantics become operationalized. When discrepancies or failures arise—such as incompatible assumptions or schema updates—agents revert to negotiation mode to revise and re-freeze the PD. Within larger networks, PDs propagate among agents through shared registries, allowing newly joined agents to adopt existing protocols without re-negotiation. This diffusion produces emergent chains of interlinked PDs (for instance, ordering, delivery, and navigation) that coordinate distributed tasks without a centralized ontology. Empirical demonstrations show that as PD reuse increases, the proportion of communications requiring LLM interpretation declines and overall cost decreases significantly, indicating that the semantic alignment established by PDs can persist and scale across heterogeneous agent networks.

\subsection{LOKA} 
\label{subsec:loka}

The LOKA Protocol \cite{rajesh_ranjan_loka_2025} proposes a decentralized framework for establishing trustworthy and ethically aligned collaboration among heterogeneous AI agents. It extends beyond conventional communication standards by embedding identity management, post-quantum security, and ethical consensus directly into the protocol layer. Unlike protocols such as MCP or A2A that primarily address interoperability and structured messaging, LOKA aims to ensure that multi-agent ecosystems can make collective decisions that are verifiable, auditable, and ethically compliant. Its design integrates four foundational layers—identity, governance, security, and consensus—while still adhering to the canonical three-layer communication taxonomy of communication, syntactic, and semantic alignment. The following subsections describe these layers in detail, emphasizing how LOKA operationalizes decentralized ethical coordination across autonomous agents.

\subsubsection{Communication Layer}

At the communication layer, LOKA ensures secure, authenticated, and tamper-resistant connectivity among agents operating across open networks. Each agent is identified through a decentralized identifier (DID) and associated verifiable credentials (VCs) that encode its provenance, capabilities, and trust level. Communication occurs through encrypted channels that employ post-quantum cryptographic primitives, notably CRYSTALS-Kyber for key encapsulation and CRYSTALS-Dilithium for digital signatures. Every interaction is digitally signed and timestamped, guaranteeing non-repudiation and traceability while remaining resistant to future quantum attacks. Agents exchange structured intent messages—such as proposals, votes, or audit logs—over HTTP or blockchain-based channels, and each message can be verified independently without reliance on a central authority. This architecture allows LOKA agents to communicate securely and maintain persistent identity relationships across heterogeneous domains, forming the infrastructural backbone for decentralized coordination.

\subsubsection{Syntactic Layer}

The syntactic layer defines how information is structured, transmitted, and validated within LOKA’s decentralized ecosystem. Each message follows a canonical JSON-based schema containing fields for agent identity, intent type, payload, timestamp, and digital signature. These schemas are immutable once published and can be extended by referencing standardized ontologies or domain-specific modules. LOKA formalizes two key message types: \textit{intent messages}, which express an agent’s proposed action or ethical stance, and \textit{decision messages}, which encode the outcomes of collective deliberation. The protocol employs a consistent lifecycle—intent issuance, encrypted voting, aggregation, and consensus recording—ensuring that every decision can be reconstructed and verified from on-chain records. Validation and error handling are machine-actionable: malformed payloads or invalid signatures automatically trigger a rejection message with standardized error codes. By enforcing schema uniformity across all agents, LOKA guarantees syntactic interoperability and reproducible reasoning traces, enabling heterogeneous agents to engage in decentralized governance without ambiguity or loss of context.

\subsubsection{Semantic Layer}

The semantic layer in LOKA governs how meaning, intent, and ethical context are aligned during multi-agent decision-making. Beyond ensuring message compatibility, it embeds moral reasoning and regulatory compliance into the consensus process through the Decentralized Ethical Consensus Protocol (DECP). Each agent maintains a \textit{Contextual Ethical Profile} (CEP) that defines its normative priorities—such as fairness, safety, or environmental responsibility—derived from human-provided ethical baselines or jurisdictional policies. During collective deliberation, agents cast encrypted votes that reflect their CEP-weighted ethical judgments. Homomorphic encryption and secure multi-party computation (MPC) allow these votes to be aggregated without exposing individual preferences, yielding a collective outcome that is both privacy-preserving and auditable. The resulting consensus is immutably recorded on a distributed ledger, along with justification metadata and verification proofs. This mechanism transforms communication from mere information exchange into a process of ethically grounded, verifiable reasoning. Through its semantic architecture, LOKA establishes a foundation for AI societies where agents can not only reach agreement but also justify decisions in alignment with shared human values.

\subsection{Agent Network Protocol} 
\label{subsec:anp}
The Agent Network Protocol (ANP)~\cite{gaowei_chang_anp_2024} aims to serve as the HTTP of the agentic internet—a general-purpose networking protocol that enables intelligent agents to discover, identify, authenticate, and communicate securely across open networks.
Unlike model-centric coordination protocols such as MCP, which focus on tool invocation within a trusted execution context, or A2A, which targets intra-organizational communication, ANP is explicitly designed for cross-domain, decentralized agent interoperability. Its architecture consists of three tightly coupled layers—communication, syntactic, and semantic—which together ensure that agents can (1) establish trust and connectivity, (2) exchange structured and interpretable messages, and (3) understand the intent and meaning behind those exchanges. Data transmission follows the HTTP standard, while information representation leverages JSON-LD (Linked Data) to embed semantic context (\texttt{@context}, \texttt{@type}) compatible with schema.org vocabularies. This combination facilitates web-native interoperability and machine interpretability at scale.

\subsubsection{Communication Layer}
At the communication layer, ANP is built on \textbf{decentralized identifiers (DIDs)} and \textbf{end-to-end encryption}, providing cryptographically verifiable identity and transport security without relying on centralized account systems.
Each agent possesses a unique DID document describing its public keys and service endpoints. The recommended implementation, \texttt{did:wba}, maps identifiers directly to a web-hosted \texttt{did.json} file, preserving decentralized control while maintaining low deployment cost and compatibility with existing HTTPS/DNS infrastructure.
A typical handshake proceeds as follows: agent A sends a signed request containing its DID to agent B; B retrieves the DID document, verifies A’s signature, and issues a time-limited access token. Subsequent interactions are token-authenticated, providing mutual accountability and auditable trust. For discovery, ANP adopts the \texttt{.well-known/agent-descriptions} convention, enabling agents to be indexed by search engines or registered locally. The overall flow—discovery → authentication → secure session → interaction—mirrors the ``\textit{can we talk?}" question of our three-layer framework, emphasizing secure connectivity and identity verifiability as prerequisites for higher-level coordination.

\subsubsection{Syntactic Layer}
The syntactic layer governs message formats and capability exposure. ANP unifies all information exchange under \textbf{JSON-LD}, extending plain JSON with linked-data semantics through \texttt{@context}. This allows agents to annotate entities, relations, and intents with globally resolvable meanings while remaining backward-compatible with existing JSON tooling. Central to this layer is the \textbf{Agent Description}, a structured manifest that includes an agent’s metadata, authentication requirements, and capability registry. Capabilities may be defined as either natural-language interfaces (for open-ended dialogue) or structured APIs (via OpenAPI, JSON-RPC, etc.).
Each description explicitly specifies callable actions, input/output schemas, and human-authorization policies, allowing agents to advertise what they can do and how they can be invoked in a machine-readable manner. Furthermore, ANP supports discovery via web-exposed listings under \texttt{.well-known/agent-descriptions}, enabling multiple agents to coexist on a single domain. This layer thus corresponds to the ``\textit{do we speak the same language?}" aspect of our survey taxonomy, standardizing message syntax and interface representation across heterogeneous systems.

\subsubsection{Semantic Layer}
The semantic layer addresses the challenge of mutual understanding.
Instead of imposing a heavyweight ontology, ANP leverages the \textbf{Linked Data} principles of JSON-LD to reference existing vocabularies (e.g., \texttt{schema:Agent}, \texttt{schema:SearchAction}, \texttt{schema:Product}). By embedding these globally interpretable identifiers, ANP ensures that entities and actions maintain consistent meaning across domains and implementations. Moreover, the protocol tightly couples identity, description, and authorization: a DID document can directly link to an agent’s description and service definitions, forming a verifiable chain from \textit{who you are} → \textit{what you claim to do} → \textit{how you are allowed to act}. This design mitigates risks such as prompt-injection or interface spoofing by ensuring semantic commitments are cryptographically grounded. Although ANP currently delegates deeper ontology alignment (e.g., domain-specific semantics or multimodal grounding) to higher-level ecosystems, its JSON-LD foundation provides a robust substrate for cross-domain comprehension, retrievability, and reasoning compatibility, which are largely absent in syntax-only protocols like MCP.

In summary, ANP provides a practical framework for open agent interoperability. Its design naturally aligns with the communication, syntactic, and semantic layers of our taxonomy: it enables decentralized and secure communication through DID-based identity, describes agent capabilities in a structured JSON-LD format, and embeds lightweight semantic links to support cross-domain understanding. However, ANP still faces several limitations. It focuses mainly on infrastructure and discovery, leaving higher-level reasoning, ontology alignment, and multi-agent coordination semantics largely undefined. As a result, while ANP offers a strong foundation for connecting agents on the open web, additional standards are needed to achieve deeper semantic interoperability and reliable large-scale collaboration.

\subsection{Coral Protocol} 
\label{subsec:coral}
Coral Protocol \cite{coral} represents an open and decentralized infrastructure designed to connect the emerging “Internet of Agents.” It provides a unified substrate for communication, coordination, trust management, and payment among heterogeneous AI agents. The protocol aims to establish interoperability across diverse ecosystems and vendors, ensuring that agents can collaborate seamlessly and securely. 
Built on modern distributed and blockchain principles, Coral adopts a layered design spanning communication, syntactic, and semantic dimensions. 
This makes it a representative example of protocols that attempt to go beyond basic interoperability toward trust-aware multi-agent coordination.

\begin{figure*}[ht]
\centering
\includegraphics[width=\textwidth]{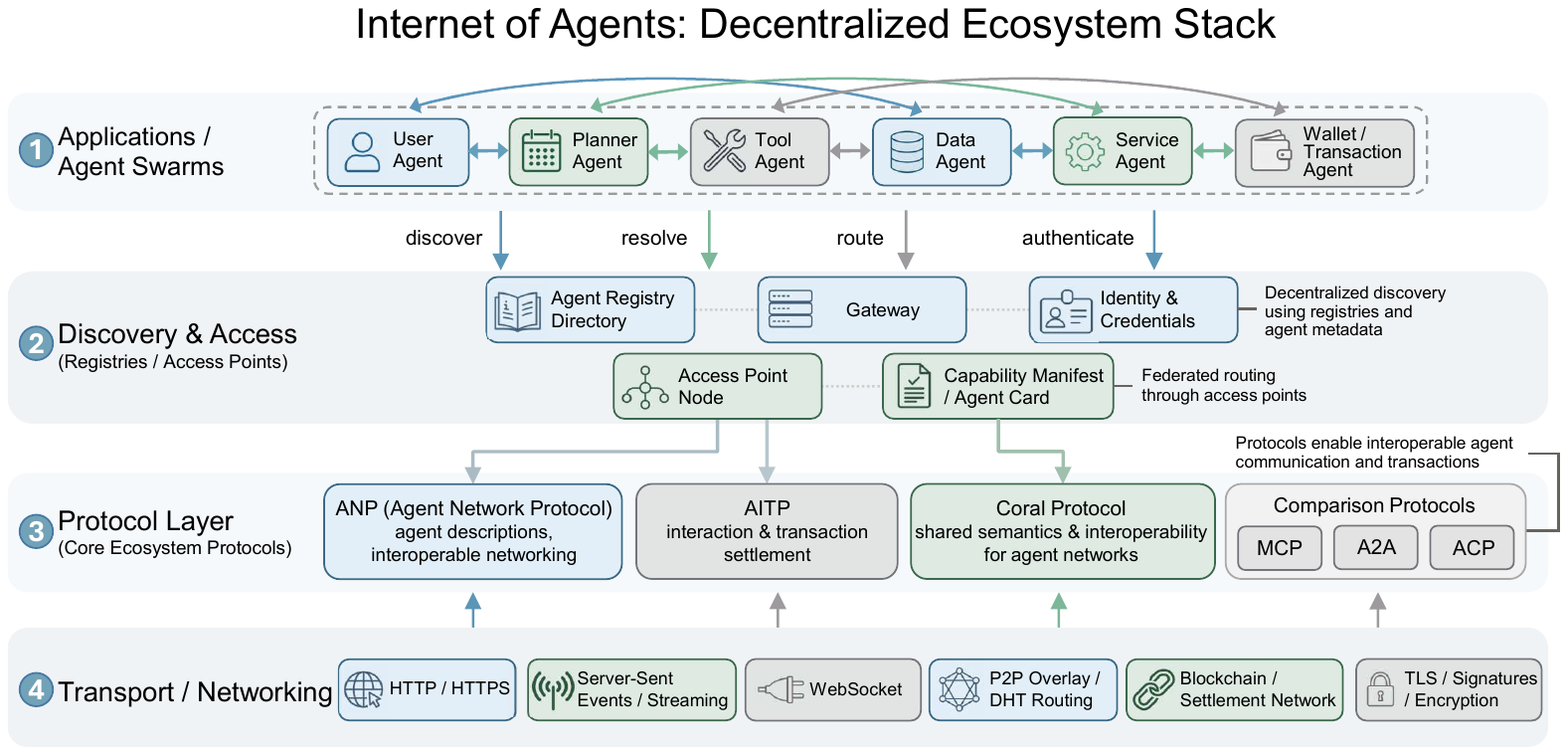}
\caption{Internet of Agents architecture. Coral Servers coordinate heterogeneous agents through decentralized message routing and task orchestration. The system supports trusted interaction and cross-agent workflows, while higher-level alignment is mainly mediated through identity, contracts, and blockchain-based verification rather than explicit clarification mechanisms.}
\label{fig:ioa-architecture}
\end{figure*}

As shown in Figure~\ref{fig:ioa-architecture}, Coral organizes agents into a decentralized network mediated by Coral Servers, enabling distributed coordination across multiple domains. From the perspective of the three-layer taxonomy, Coral provides strong support for communication and syntactic structuring, while its semantic mechanisms focus more on identity, trust, and verification than on interactive clarification. In particular, clarification is largely absent, and context alignment is achieved through shared infrastructure, contracts, and verifiable records rather than explicit conversational repair.

\subsubsection{Communication Layer}
The communication foundation of Coral is organized through a network of \textbf{Coral Servers}, each serving as a hub that hosts Coralised Agents, Model Context Protocol servers, developer tools, and wallets. These nodes interconnect over the Internet to form a distributed and secure environment for multi-agent collaboration. The transport layer is intentionally flexible and Internet-native, allowing agents to communicate over standard protocols while benefiting from end-to-end authentication and encrypted message routing.
Each Coral Server manages message delivery and task orchestration among Coralised Agents. When a user submits a query through an application, it is dispatched by the Coral Server’s Interaction Mediation component to the relevant agents. Coral thus generalizes the concept of transport-level messaging into a structured multi-agent coordination system, handling discovery, sub-task allocation, and response aggregation. Underlying this process is the use of the Model Context Protocol, which standardizes access to models, tools, and external computation backends. In this architecture, Coral focuses on inter-agent communication, while MCP manages model and tool connectivity, jointly enabling scalable distributed reasoning.
This design ensures that the communication layer does not simply deliver data packets but provides verifiable interaction primitives including authenticated exchanges, task routing, and compositional workflows between agents hosted across the global Coral network.

\subsubsection{Syntactic Layer}
At the syntactic level, Coral defines the interface contracts and invocation semantics that make agent interaction uniform and composable. Any existing AI model, tool, or service can be ``\textbf{Coralised}", meaning it is wrapped with a protocol-conformant interface exposing its capabilities, metadata, and invocation schema. This Coralisation process enforces standardized request–response formats and typed envelopes for communication, error handling, and artifact exchange, so that heterogeneous systems become interoperable components of a shared ecosystem. The syntactic framework of Coral also coordinates closely with MCP. While MCP provides the structured resource and tool interface (for example, through OpenAPI-like schemas), Coral organizes those resources within higher-level agent workflows. Through the Coral Server’s Interaction Mediation, messages between agents are enriched with metadata about task context, delegation, and response linkage, enabling multi-step reasoning chains that can span multiple servers and organizations. This layer effectively transforms a network of isolated APIs into a unified multi-agent operating environment, in which invocation, composition, and discovery are conducted through consistent message schemas.

\subsubsection{Semantic Layer}
The semantic dimension of Coral addresses meaning, trust, and value exchange. It introduces mechanisms for \textbf{identity management}, \textbf{secure collaboration}, and \textbf{economic coordination}, ensuring that multi-agent workflows are not only functional but also verifiable and trustworthy. Each agent is assigned a \textbf{decentralized identifier (DID)} linked to cryptographic credentials, allowing authentication and provenance tracking without centralized authority. Agents can form teams through verifiable contracts specifying their roles, permissions, and task-level agreements. These contracts can be registered and enforced on-chain, providing tamper-evident audit trails and governance of multi-agent collaboration. Coral extends these semantic assurances into the economic layer through \textbf{blockchain-integrated payment and reputation systems}. A built-in escrow mechanism, implemented via smart contracts on Solana, enables automatic, trustless compensation for completed tasks. Payments are held in a contract-controlled vault and released based on fulfillment criteria recorded on-chain. Each transaction is cryptographically signed, timestamped, and publicly auditable, forming a transparent trust ledger. This allows agents to transact safely even in open, cross-domain collaborations. Beyond payments, Coral’s semantic infrastructure also supports \textbf{reputation-based coordination}, where each agent’s history and reliability inform future team selection. This creates an emergent, self-regulating trust ecosystem, complementing the purely technical interoperability of the communication and syntactic layers.

In summary, Coral protocol advances the state of multi-agent interoperability by extending the conversation beyond connectivity and format alignment toward verifiable meaning, trust, and incentive alignment. Its layered design: communication for reliable and authenticated exchange, syntactic for uniform representation and invocation, and semantic for identity, reputation, and payments, embodies the transition from ``\textit{can agents talk?}" to ``\textit{can they trust, understand, and transact?}" within an open and vendor-neutral Internet of Agents

\subsection{Agent Interaction \& Transaction Protocol} 
\label{subsec:aitp}
AITP \cite{near_aitp_2025} proposes a standard for agent-to-agent and user-to-agent communication across trust boundaries, pairing a chat-thread–centric core with an extensible capabilities system that goes beyond plain text to include structured UI, forms, payments, and human-in-the-loop attestations. It aims to be for agent interactions what HTTP and HTML are for the web, an application-layer substrate that multiple runtimes and frameworks can target, while remaining complementary to orchestration frameworks that handle intra-organizational workflows and service-metadata protocols such as MCP. The current design defines three core concepts: \textbf{Threads}, which represent units of dialogue and work; \textbf{Transports}, which carry thread messages; and \textbf{Capabilities}, which agents negotiate to enable richer, typed exchanges. Specifically, Transports ensure that agents can communicate at all, threads and message contracts define a common syntax for communication, and capability negotiation with intent-bearing payloads provides the semantic grounding needed for mutual understanding. This makes AITP a representative example of a protocol that explicitly integrates both interaction and transaction across agent boundaries.

\subsubsection{Communication Layer}
At the communication level, AITP adopts the \textbf{thread} as its primary session abstraction and allows multiple transport mechanisms, typically HTTP/S with streaming primitives, to carry thread messages. The design emphasizes portability so that the same thread semantics can operate across channels while preserving ordering and progressive delivery for interactive experiences. Its model is inspired by, and compatible with, the OpenAI Assistants/Threads API, simplifying integration with existing agent stacks. AITP explicitly addresses the challenge of crossing trust boundaries, such as when a user-owned personal assistant communicates with a service- or business-owned agent. These interactions require clear identity declarations, negotiated capabilities, and explicit authorization steps before sensitive operations like payments or attestations proceed. The protocol’s operational model supports both long-running and progressive exchanges, while keeping transports modular to allow either stateless calls or fully interactive streaming modes.

\subsubsection{Syntactic Layer}
At the syntactic level, AITP defines a uniform \textbf{message} schema organized around threads, consisting of conversational turns, attachments, and machine-actionable status markers. Because its design aligns with popular assistant APIs, client and server roles are well defined: clients submit inputs, agents respond with text or structured payloads, and both can append or stream updates until completion. The capability system extends this schema by defining typed message families that structure non-text interactions such as generative UI, forms, payments, attestations, and multimodal input. During capability exchange, agents advertise supported message types so that peers can tailor outputs appropriately, e.g., returning a structured payment object rather than plain text if the peer supports the payment capability. The specification lists standard capability schemas to ensure validation and graceful degradation. Importantly, AITP is designed to complement other standards rather than replace them: service-oriented agents may still use MCP or internal orchestration frameworks like AutoGen or LangGraph, while exposing an AITP-facing interface for inter-agent communication. This helps bridge heterogeneous ecosystems without imposing new infrastructural constraints.

\subsubsection{Semantic Layer}
Semantically, AITP enables intentful interactions through \textbf{capability} payloads that encode the meaning and requirements of each transaction. A form capability conveys expected fields and constraints; a payment capability specifies payee, amount, and authorization logic; and an attestation capability captures human approval or oversight. By exchanging typed intents rather than unstructured text, agents can align on both the purpose of an action and the evidence or consent it requires, improving safety and reliability when crossing organizational or trust boundaries. The protocol intentionally avoids mandating a global ontology. Instead, it standardizes a family of interaction schemas and allows peers to negotiate which ones they support, achieving semantic alignment by contract while retaining domain-level flexibility. For deployments requiring deeper semantic grounding, such as identity, units, or compliance invariants, AITP can be composed with higher-level policies described in the earlier Semantic Layer framework. Because transactions often imply authority and liability, AITP’s semantics explicitly represent who is acting for whom, what capability is being exercised, and what confirmations are required, supporting auditability and mitigating the ``\textit{syntax-only}" failure mode of previous designs.

Overall, AITP combines a threaded conversation core with capability-typed interactions to enable secure, structured, and extensible cooperation between independently owned agents. It embodies the three-layer framework proposed earlier: minimal but robust communication transports, precise message and capability schemas, and intent-bearing semantics that ensure alignment and safety, while coexisting naturally with orchestration and service protocols already in use across the agent ecosystem.
\newcommand{\cyk}[1]{{\color{blue}{[CYK: #1]}}}

\subsection{Eclipse Language Model Operating System} 
\label{subsec:lmos}
Eclipse Language Model Operating System (LMOS) protocol represents a full-stack architecture for establishing a vendor-neutral ``Internet of Agents''~\cite{eclipse_language_2025}.
Unlike protocols narrowly for tool invocation, LMOS provides a holistic framework for agent discovery, description, communication, and semantic interoperability. 
It is explicitly designed for ``big world'' use cases, enabling agents and tools from different organizations to interconnect, regardless of their underlying technologies.
To achieve this, LMOS builds upon the W3C Web of Things (WoT), adapting its principles of protocol-agnostic abstraction and semantic description to the domain of multi-agent systems.

\subsubsection{Communication Layer}
At the communication layer, LMOS's primary design choice is \textbf{transport-protocol agnosticism}. 
It does not mandate a single transport mechanism but instead leverages the W3C WoT protocol binding abstraction. 
This allows an agent's abstract interaction model, defined at the syntactic layer, to be mapped to concrete transport protocols, e.g., HTTP, WebSockets, MQTT, etc.

This abstraction provides immense flexibility, but LMOS also specifies robust mechanisms for security and discovery. 
Specifically, security and identity are cryptographically grounded using W3C DIDs, which provide a verifiable and decentralized method for agent authentication. 
As for the discovery design, LMOS defines a multi-faceted architecture that supports (1) local network discovery via DNS-SD/mDNS for zero-configuration environments and (2) global discovery via federated agent registration. 
This robust discovery mechanism ensures that agents can reliably find and hear each other across heterogeneous network environments, fulfilling the primary role of the communication layer.

\subsubsection{Syntactic Layer}
The syntactic layer of LMOS is where its unique design becomes most apparent. 
The protocol standardizes on \textbf{JSON-LD} as its core data format for all agent and tool metadata. 
Furthermore, LMOS specifies a standardized ``Agent and Tool Description Format''. 
Expressed in JSON-LD, this manifest defines an agent's capabilities, metadata, and interaction affordances. 
Crucially, this description document must be cryptographically signed using the agent's DID.
This signature guarantees both the authenticity, \textit{i.e., proof of origin}, and integrity, \textit{i.e., proof against tampering}, of the capability description, a critical feature for open-network interoperability.

The syntactic layer is tightly bound to the communication layer, via the W3C WoT \textit{Thing Description (TD)}, and bridges the connection to the semantic layer.
Specifically, the TD document syntactically defines the technical instructions that explicitly detail the protocol bindings, such as the precise endpoint URLs, HTTP methods, or WebSocket topic, that are required to invoke a specific action.
This mechanism effectively translates the semantic goal, e.g., \textit{invoke action}, into a well-defined and machine-executable instruction for the communication layer.

\subsubsection{Semantic Layer}
While many protocols remain lightweight at the semantic layer, LMOS is explicitly designed to solve semantic interoperability.
It achieves this target by three primary mechanisms. 
Specifically, the protocol's use of JSON-LD is fundamentally a semantic choice. 
It allows agent metadata to link to standardized, globally resolvable vocabularies and ontologies. 
This provides shared meaning beyond ambiguous natural-language strings, e.g., linking \texttt{book\_flight} to a formal travel ontology.
Second, LMOS adopts the W3C WoT's \textbf{Abstraction Interaction Model}, standardizing all interactions around three semantic primitives: \textit{Properties (stateful data that can be read/written)}, \textit{Actions (operations that can be invoked)}, and \textit{Events (asynchronous messages that can be subscribed to)}. 
This standardized model of interaction provides a shared conceptual grounding for all agent behaviors. 
Finally, LMOS proposes a ``meta-protocol layer'' (or a ``protocol of protocols''). 
This advanced concept allows agents to dynamically negotiate the optimal communication protocol for their specific task. 
This negotiation itself is a form of deep semantic alignment, where agents collaboratively determin the optimal language for their task, rather than being restricted to a single predefined one. 
In summary, LMOS's semantic layer provides a rich framework for shared meaning that far exceeds the syntactic-level guarantees of protocols like MCP.





\subsection{Spatial Population Protocols}
\label{subsec:spp}
The SPP represents a novel theoretical model of distributed computing for robot agents.
This model extends classical population protocols by embedding agents in a Euclidean space and granting them the ability to perform ``geometric queries'' during their interactions. 
The primary application discussed is solving the Distributed Localisation Problem, where anonymous agents must converge on a unified coordinate system from arbitrary initial states.
SPP primarily focuses on the algorithmic innovation in the semantic layer.



\subsubsection{Semantic Layer}
The Semantic Layer of the SPP model is its core component, as it explicitly defines the protocol's ultimate semantic goal.
The SPP's explicit purpose is to solve the Distributed Localisation Problem. 
Thus, the protocol's goal is to make all agents reach a ``self-stabilizing'' consensus on their coordinates. 
A unique feature of SPP is that it uses a precise coordinate update algorithm to force this consensus. 
Specifically, the mechanism of this algorithm works as follows.
First, the protocol establishes a subset of agents whose coordinate labels are already mutually aligned, meaning every agent in this subset shares the same offset between its label and its true position. Second, whenever an agent with an inconsistent coordinate interacts with a already-consistent agent, the update rule forces the inconsistent agent to adopt the consistent coordinate system, thereby joining the aligned subset.
SPP proves that this correct coordinate system spreads through the population like a one-way epidemic. Once all agents have joined the aligned subset, no further updates are performed and the protocol stabilises silently. At this point, all agents' coordinate labels are stable and perfectly consistent relative to each other, which eventually solves the Distributed Localisation Problem.

\subsection{Predict \& Explain Protocol} 
\label{subsec:pxp}
The PXP directly addresses the challenge of ``two-way intelligibility'' in multi-turn interactions human-LLM interactions~\cite{ashwin_srinivasan_implementation_2024}.
It is a structured interaction model designed to mediate interactions for collaborative data analysis tasks.
The system models the interaction as two communicating finite-state machines: a ``human-agent'' and an ``LLM-based agent''. 
The protocol's primary goal is to move beyond simple natural language prompting and provide a formal mechanism to capture and measure the degree of mutual understanding.

\subsubsection{Communication Layer}
The communication layer is realized as a rudimentary blackboard system, where agents do not communicate directly~\cite{ashwin_srinivasan_implementation_2024}.
Instead, a central \texttt{INTERACT} procedure acts as a scheduler, alternating between two agents.
``Signal transmission'' occurs when an agent is called, generated a message, and writes that message to a shared relational database. The \texttt{INTERACT} procedure then reads this message from the blackboard and uses it as context for polling the other agent. 
The database itself is the communication channel that acts as a shared and asynchronous message buffer.

\subsubsection{Syntactic Layer}
The syntactic layer is the formal core of the PXP. It defines a rigid message format that all agents must follow: every message consists of three components: a tag, a prediction, and a natural language explanation for that prediction. The tag must be drawn from a fixed vocabulary of five options: INIT, RATIFY, REFUTE, REVISE, and REJECT.
The protocol's transition function, defined by the AGENT procedure, formally governs this syntax. It requires each agent to compare its own internal prediction and explanation against those of the previous message, using two dedicated comparison functions. The outcome of these comparisons deterministically dictates which tag the agent must assign to its response.

\subsubsection{Semantic Layer}
The semantic layer is the entire purpose of PXP.
The protocol's semantic goal is to determine if the agents truly understand each other's reasoning, not just their final answer.
The above four tags are the explicit carriers of semantic meaning.
Specifically, \texttt{RATIFY} suggests that two agents semantically align both the prediction and explanations with each other.
\texttt{REVISE} means that the explanations of an agent $a$ is intelligible and semantically persuasive. 
Another agent $b$ will updating my internal model based on $a$'s reasoning.
\texttt{REFUTE} means that the agent $a$ understands $b$'s reasoning, but finds it it semantically incorrect or insufficient, and $a$ will not update the model.
\texttt{REJECT} indicates that two agents have a fundamental semantic misalignment on both the prediction and the explanation.
The protocol measures semantic success by analyzing the sequence of these tags over a session.
For example, a session is deemed ``One-Way Intelligible'' if it contains \texttt{RATIF} or \texttt{REVISE} tags and, crucially, no \texttt{REJECT} tags, signifying a successful exchange of meaning without a fundamental breakdown.

\subsection{Web-Agent Protocol} 
\label{subsec:wap}
The WAP is an open standard and framework for collecting, interpreting, and replaying real user interactions on the web~\cite{ota_tech_ai_web_2025}.
Its primary purpose is to act as a bridge to ground LLM-based agents in real-world web environments.
It does this by defining a standard for capturing human demonstrations as structured and replayable artifacts.
Such record and replay model is fundamentally asynchronous that is designed to create executable and MCP-compatible agent services from human demonstrations.

\subsubsection{Communication Layer}
WAP's communication layer mainly address its two distinct operational modes, i.e., \textit{Recording} and \textit{Replaying}.
In WAP, Recording component is a Chrome extension.
This extension captures user interactions and transmits them to a dedicated server. 
This server listens on a standard web endpoint, e.g., http://localhost:4934/action-data, which confirms a client-server transport model for data collection phase.
As for Replaying component, WAP delegates execution to MCP.
The framework converts a recorded task trace into a brand-new and standalone MCP server.
An LLM-based agent can then communicate with this newly generated WAP service using standard MCP transport mechanisms.
Therefore, WAP acts as a generator of MCP-compatible services.

\subsubsection{Syntactic Layer}
The syntactic layer is the core contribution of WAP, the protocol of which is the standardized data format for the task trace in two forms:
(1) \textit{Raw Event Stream}.
This is the initial JSON format captured by the Chrome extension.
It contains low-level and detailed information, including \texttt{taskID}, \texttt{type}, \texttt{eventTarget} and full \texttt{pageHTMLContent}~\cite{ota_tech_ai_web_2025}.
(2) \textit{Replay List}. The raw event stream is processed by scripts into a final and high-level JSON replay list.
This replay list artifact is the definitive syntax that the newly generated MCP server uses to define its capabilities. 
It replaces screen scapring with a robust and machine-readable scheme that captures information such as \texttt{DOM context}, \texttt{page content}, and \texttt{timestamps}.

\subsubsection{Semantic Layer}
WAP's semantic layer is designed to capture action-level intent and goal-oriented steps, which enables generalization beyond the original recording. 
This semantic richness is achieved by processing the raw event stream into two distinct semantic modes: \textit{Exact Replay} and \textit{Smart Replay}.
In Exact Replay mode, the agent follows the trace precisely as a script that the semantic goal prioritizes perfect fidelity over adaptability.
Smart Replay mode processes the raw events to produce condesned goal-oriented steps.
This process may uses LLMs to infer the user's subgoals from the low-level actions. 
The resulting artifact allows a replaying LLM-based agent to understand the user's goal and adaptively attempt to complete it, even if the website's DOM structure has changed since the recording.



\section{Technical Debt in Current Agentic Protocol} 
\label{sec:techdebt}

Following our comprehensive analysis of the foundational communication paradigms of existing protocols, we must now confront the long-term viability and hidden costs of these designs. The rapid development of agentic communication protocols often necessitates trade-offs in design simplicity, documentation, and standardization, inevitably resulting in technical debt. This debt manifests as accumulated complexity that hinders maintenance, slows future evolution, and increases the risk of subtle bugs. In this section, we will therefore shift our focus to identifying and analyzing technical debt in agent protocols from three layers.


\subsection{Ecosystem Fragmentation Debt}
The ecosystem of Multi-Agent Systems (MAS) and agent coordination is currently experiencing significant fragmentation. Multiple  partly overlapping protocols coexist in the wild, each with its own design pacing, assumptions and target scenarios. This heterogeneity can be justified: MCP focuses strongly on tool invocation and context sharing, ACP deals with multi-modal messaging and session‐based agent coordination, A2A emphasizes peer-to-peer agent delegation via “Agent Cards”, and ANP is oriented toward open-network discovery with decentralized identifiers (DIDs). Nonetheless, from the viewpoint of a researcher or enterprise architect, this divergence poses serious risks to long-term interoperability, maintainability and ecosystem scalability.

\subsubsection{Custom Integration Cost}
When an organization adopts more than one of these protocols (for example MCP for tool integration and A2A for broader agent collaboration), engineering teams are required to build translation layers, glue-code and adapters that reconcile differences in transport (JSON-RPC vs REST vs streaming), discovery mechanism, identity/authentication model, data schema, and life-cycle management. The survey literature explicitly describes this as an emerging “integration debt”. 

In effect, each additional protocol supported multiplies the integration surface area. Every change in one protocol version or downstream component may trigger cascading adjustments in adapters or custom code. Over time, this accumulates what might more properly be termed “interoperability debt” rather than mere technical debt: the latent cost of supporting multiple, diverging protocol stacks.

\subsubsection{Ecosystem Duplication and Lock-in}
Another consequence of protocol divergence is duplication of effort and potential vendor lock-in. When vendors or platforms choose a specific protocol stack, they build tool-chains, agent registries, and services around it. If a partner or external agent uses a different stack, bridging is required—or the organization must re-implement across stacks. This reduces capability and increases cost. The literature suggests that until protocols converge or robust translation layers exist, agent ecosystems risk becoming “vertically integrated stacks” that do not interoperate fluidly. 
From a scientific viewpoint this means that the promise of “a web of agents” or “agent economy” becomes less likely unless interoperability debt is actively managed.

\subsubsection{Governance and Strategic Uncertainty}
Beyond direct engineering cost, there is a governance and strategic dimension: the absence of a unified protocol stack means organizations face uncertainty in their roadmap planning for agent ecosystems. Which protocol should be adopted and invested in? Will one become dominant, or will multiple continue to coexist indefinitely? Are tools built on MCP still interoperable when agents built on A2A must collaborate? The survey emphasizes that choices made now may lead to lock-in or stranded systems. 


\subsection{Technical Debt in the Communication Layer}
The newest wave of Agent-to-Agent (A2A) communication protocols represents a necessary architectural pivot away from the complexity of legacy standards. These protocols prioritize operational efficiency, adopting lightweight structures like JSON-RPC 2.0 over HTTPS to align with modern cloud and web-native architectures.
However, this shift has introduced a new technical debt TD focused on the communication infrastructure itself. This debt stems from the omission of standardized, robust features for complex coordination and failure analysis, resulting in the following issues.

\subsubsection{Incomplete Session Management}
Despite advances in agent communication protocols, such as structured lifecycles and explicit session identifiers, coordination complexity remains. Frameworks like A2A and MCP clarify session persistence and handoff but still assume stable, synchronous interactions. In practice, MAS systems are fluid: agents join or leave dynamically, act asynchronously, and pursue overlapping or conflicting goals. This exposes a persistent “session management debt,” where rigid lifecycles and static state transitions fail to reflect the temporal and behavioral diversity of large, heterogeneous coordination environments.

\textbf{State-model mismatch for long-running, asynchronous dialogues}: Protocol lifecycles that assume short, well-bounded phases (e.g., Initialization → Operation → Shutdown) simplify implementation but poorly model dialogues that: (a) persist for minutes–hours with intermittent participant availability, (b) require speculative parallel negotiation and later reconciliation, or (c) perform asynchronous handoffs between agents with different capabilities. Empirical and formal surveys note that session models which assume synchronous progress or well-formed short sessions struggle to represent these patterns. 

\textbf{Partial failures and uncertainty}: Long-running sessions expose partial failure modes (crash, network partitions, delayed messages) that break assumptions implicit in many session specifications. Recent work on asynchronous multiparty session types explicitly identifies the need to model crash-stop and unreliable delivery semantics to retain safety guarantees.

\textbf{Interleaving and hand-off complexity}: Real MAS interactions often interleave multiple logical sessions (delegation, nested sub-protocols). This complicates local reasoning and can reintroduce deadlocks or protocol mismatches unless the protocol model supports composition and dynamic endpoint creation. The MPST literature highlights this as a major source of coordination debt.

\subsubsection{Inadequate Error Management}
Modern A2A protocols leverage standard HTTP status codes (e.g., 200, 404, 500) and structured JSON-RPC error objects to report communication failures. While effective for handling network and operational errors (like a service being unavailable or a resource not being found), this model fails to communicate agent-specific, policy-driven failure states, leading to ambiguity and high application-level error handling complexity

\textbf{The deficiency of generic error handling}: While standard HTTP status codes are effective for basic transport, network, or server‐side operational errors, they accumulate operational errors. For example, suppose an agent receives a request that violates its internal policy or it must “refuse” a request, or it fails because it cannot “understand” a performative or content. These are cognitive or policy-driven failures—not merely transport or protocol faults. Relying solely on generic HTTP status codes obscures this distinction. The recipient agent cannot easily tell whether the remote agent simply timed out, encountered a network error, or actively rejected the interaction for policy reasons.

\textbf{The burden of custom error routines}: Because the protocol layer only provides generic or insufficient error signals, developers are forced to implement elaborate application-layer error‐handling logic to interpret the true nature of breakdowns. This means building custom logging, diagnostic heuristics, retry logic, fall-back behaviors, and agent‐specific exception handling. In other words, the reliance on generic transport‐level error codes shifts complexity into the application layer, increasing operational overhead and reducing system robustness. This is especially problematic in multi‐agent settings where one agent’s inability to interpret another’s refusal can cause mis‐delegation or task starvation.

\subsubsection{Inadequate Message Structure}
In conventional networking protocols, message formats are rigorously defined with fixed headers, explicit fields, and precise semantics for routing, error signalling, and sequencing. Agent-to-agent messages focus on semantic content such as intent and context but often lack structural rigor, versioning, and standardized error semantics, creating technical debt that increases complexity and reduces interoperability.

\textbf{Insufficient rigor and field definitions}: Many agent-protocol messages establish only the most minimal structural framework (for example, the basic role, parts, and metadata elements found in A2A), but they fail to rigorously define the use or composition of their core components. This means that large sections of the message structure remain vague or optional, leading to scenarios where different agents interpret fundamental fields, such as "metadata" or "parts," in divergent ways. Because the required message formats are so loosely defined, agents cannot rely on standardized parsing. Instead, they must incorporate significant extra, customized logic simply to interpret the structures they receive or convert them into a usable internal format. This necessity for complex, bespoke processing substantially increases the overall system complexity and, simultaneously, reduces the reliability of communication across the agent network.

\textbf{Versioning, compatibility, and extensibility issues}: Another major challenge stems from inadequate provisions for versioning, extensibility, and compatibility. Networking protocols typically incorporate explicit version numbers, reserved fields, and well-defined extension mechanisms that ensure backward compatibility and facilitate evolution across networked systems. By contrast, many A2A protocols lack such formal mechanisms, creating the risk of incompatibility as agent implementations evolve or interact across different platforms. Even protocols such as the Agent Network Protocol (ANP), which introduce structured binary headers with version information, have yet to achieve widespread adoption. As a result, agents operating in heterogeneous environments may encounter messages that they cannot parse or interpret correctly, requiring custom adapters or bridging logic.

\textbf{Messaging Resilience Deficiencies}: While standard networking messages typically incorporate essential meta-fields for reliability—including sequencing, ordering, acknowledgment (ACK/NACK), timeouts, and retries, agent messages seldom include these constructs inherently. Instead, agent protocols typically rely on the underlying transport layer (such as HTTP or WebSockets) for basic reliability and ordering, rather than managing these functions at the message level. This reliance creates ambiguity: the critical semantics of "was this message successfully delivered and processed?" remain unclear within the agent message itself. Consequently, agents cannot seamlessly or reliably differentiate crucial failure states like Message Loss , Refusal/Rejection, or Message Staleness.

\subsection{Technical Debt in the Syntactic and Semantic Layers}
This subsection shifts the focus to accumulated technical debt arising from design decisions that prioritize operational flexibility and convenience over machine-enforceable rigor and unambiguous meaning. Unlike the debt observed in the Communication Layer, which relates to transport reliability and session state, syntactic and semantic debt manifests as fundamental failures in data integrity, message interpretation, and the long-term coherence of agent interaction.

\subsubsection{Lack of Syntactic Rigor}
Syntactic debt in agent protocols stems from the lack of formal, machine-readable specifications for the message content payload itself, separate from the basic message envelope. While the previous section noted the absence of generic resilience fields (e.g., sequencing), this subsection discusses the critical failure to define the internal data structure being exchanged. This deficiency introduces significant costs related to data model management and parsing fragility.

\textbf{Data Model Debt}:
The modern preference for lightweight, JSON-based messaging without strict schema validation or Interface Definition Language (IDL) enforcement creates significant Data Model Debt, echoing issues in real-time data streaming. Most protocols define only minimal envelope structures, and leave message content (such as domain data, tool outputs, or context objects) weakly typed and inconsistently formatted. Without standardized specifications, agents lack reliable parsing guarantees.

\textbf{Schema Drift Debt}: This weak structural enforcement further introduces Schema Drift, where data types evolve silently over time. For example, a $user\_id$ field may shift from an integer to a string. Protocols using dynamic LLM outputs are especially vulnerable. Such changes seldom trigger transport-level errors but instead corrupt internal state or generate invalid downstream inputs.

These syntactic faults often lead to semantic failures. When malformed data reaches an agent or LLM, the model may reinterpret it contextually, producing hallucinations or refusals that obscure the root cause. Consequently, a simple format error can escalate into non-deterministic semantic breakdowns, complicating debugging and extending recovery times.

\subsubsection{Semantic Ambiguity Debt}
Semantic debt represents a more subtle form of technical debt in modern agentic systems. It arises from the fundamental difficulty in ensuring consistent and shared meaning across autonomous, stochastic components.

\textbf{Performative Ambiguity and Stochastic Intents}:
The industry shift away from traditional, formal protocols to lightweight, LLM-driven communication protocols introduces non-determinism into the core act of communication: determining intent. Classic protocols relied on formally specified performatives (speech acts like request, inform, or reject) and rigorous conversation semantics to establish clear, unambiguous intent, providing a predictable sequence of interactions.

However, LLM agents introduce inherent stochasticity. While they can generate structured outputs denoting intent, they may misclassify performatives or produce responses with ambiguous intent due to contextual nuances, unclear prompts, or reasoning flaws. This unreliability in signaling intent constitutes Performative Ambiguity Debt. Misclassifying a proposal as an inform, for instance, leads to invalid state transitions, miscoordination, or deadlocks that prevent goal completion. Protocols must therefore mitigate this non-determinism through strict parsing, fallback mechanisms, and deterministic validation—often requiring extensive custom engineering.

\textbf{Ontological Grounding Debt and Domain Misalignment}:
Beyond the performative’s intent, message content must also be interpreted consistently. The lack of shared knowledge frameworks creates Ontological Grounding Debt. In classic networking and communication protocols, formal ontologies (e.g., controlled vocabularies and domain models) ensured consistent interpretation of complex data, enabling coherent reasoning and collective awareness.

Existing agent communication protocols often omit or heavily simplify this layer, leaving agents to ground meaning independently through stochastic inference, internal vector stores, or RAG mechanisms. This independence leads to domain misalignment when agents apply different conceptual models to identical facts (e.g., defining “high-risk” or “non-performing loan”). Without a shared semantic foundation, coordinated reasoning and cross-organizational collaboration degrade as translation and reconciliation costs increase.

\textbf{Contextual Drift Among Agents}:
For long-horizon, multi-step tasks, maintaining semantic consistency across extended dialogues introduces a form of debt tied to the context window limitations of LLMs. In multi-agent systems, conversations span multiple agents, each with its own internal context. While an individual agent may maintain coherence through external memory or context engineering, the protocol itself cannot guarantee consistent interpretation across agents. This lack of shared context enforcement constitutes Contextual Coherence Debt at the communication protocol level.

Mitigating this debt requires continuous Context Engineering on each agent, including compaction (summarizing near-limit dialogues), structured note-taking (persisting distilled knowledge externally), and coordinated retrieval mechanisms. However, even with robust agent-side strategies, misalignment can occur if contexts diverge across participants. This divergence (i.e., Contextual Drift among agents) undermines shared situational awareness, causing mis-coordination, inconsistent decision-making, and costly re-synchronization. The debt exists because the protocol itself does not provide mechanisms to enforce or synchronize multi-agent context, forcing extensive engineering effort to maintain reliable coordination.

\section{What are my go-to protocol options?} 
\label{sec:go-to-option}

This section converts the survey into a binary-branch selection workflow. Beginning with deployment control and tenancy, the path narrows through intra-organizational orchestration needs, open-web discovery and identity, repetition that justifies routine extraction, trust-boundary transactions, demonstration-grounded capabilities, and governance. Each branch terminates at a concrete protocol introduced earlier (\S\ref{sec:protocol}). The organizing principle is to choose the minimal protocol whose native guarantees reduce the specific debts highlighted in the survey, including drift in session state, drift in schema, and drift in meaning, without introducing unnecessary machinery.

\subsection{Do I control both ends, and is the system single-tenant?}
\label{sec:single-tenant}
In single-operator deployments where the primary need is reliable tool invocation with strong message contracts, the natural default is \textbf{MCP} (\S\ref{subsec:mcp}). The protocol supplies low-friction transport through stdio or streamable HTTP and a JSON-RPC message contract. Addresses the communication and syntactic layers while remaining intentionally light on semantics. This design enables interoperability across components without forcing a shared ontology.

Recurring multi-step API routines in the same setting benefit from layering \textbf{agents.json} on top of MCP (\S\ref{subsec:agents-json}). Declaring actions, flows, and links over existing OpenAPI surfaces transforms static API descriptions into executable multi-step manifests that an agent can call deterministically. This approach remains lightweight and does not require a formal semantic layer.

MCP works best when both the agent runtime and its tool interfaces operate within one administrative domain, whether or not they are literally owned by the same team. When interfaces evolve independently or span different organizations with separate governance, coordination, or version control, the environment becomes multi-party. In such cases, the workflow proceeds to Section~\ref{sec:intra-org}, where orchestration protocols such as ACP introduce discovery, session persistence, and schema negotiation.

\subsection{Inside one organization: do you need streaming runs, nested calls, and durable session context?}
\label{sec:intra-org}
Service-mesh environments that involve long-running tasks, nested tool or agent calls, and incremental outputs align with \textbf{ACP-AGNTCY} (\S\ref{subsec:acp-agntcy}). Its evented HTTP/SSE model and typed event taxonomy, which covers lifecycle, message, tool-call, and state, provide uniform orchestration semantics. Threads preserve session context across agents, and OASF-style capability descriptors allow machine-readable discovery and composition.

Teams that prefer a simpler schema based on roles and multimodal Parts, while retaining streaming and optional routing for multi-agent workflows, can use \textbf{ACP-IBM} (\S\ref{subsec:acp-ibm}). Agents remain ordinary web services using SSE, and a minimal manifest combined with an optional router coordinates collaboration efficiently with limited overhead.

When neither streaming nor durable threading is required, remaining on \textbf{MCP (+ agents.json)} internally is sufficient and avoids unnecessary orchestration complexity (\S\ref{subsec:mcp}, \S\ref{subsec:agents-json}).

\subsection{Cross-organization on the open web: is discovery and delegation first-class?}
The Agent-to-agent collaboration between organizations is effectively supported by \textbf{A2A} (\S\ref{subsec:a2a}). Agent Cards enable structured discovery, and HTTP/SSE provides live streaming sessions. The protocol defines task lifecycles and clarification messages that make delegation recoverable and auditable while avoiding the need for a heavy shared ontology.

Cross-organizational collaboration that also requires verifiable identity and portable linked semantics extends naturally to \textbf{ANP} and \textbf{LMOS} (\S\ref{subsec:anp}, \S\ref{subsec:lmos}), where ecosystem alignment is advantageous. JSON-LD descriptions bound to decentralized identifiers carry resolvable meaning across domains. This approach reduces interpretation risk beyond simple message syntax and enables trust-preserving federation.

\subsection{Do you expect repeated exchanges where cost and ambiguity must drop over time?}
Interaction patterns that begin in natural language but later recur with stable intent correspond to \textbf{Agora} (\S\ref{subsec:agora}). Negotiation converges into a hashed Protocol Document, and subsequent requests follow deterministic routines that reference this document. The structure reduces repeated computation while maintaining a clear path back to renegotiation when drift occurs.

\subsection{Do actions cross trust boundaries and require typed transactions, attestations, or payments?}
Cross-ownership exchanges that need to move beyond untyped text to verifiable and auditable interactions align with \textbf{AITP} (\S\ref{subsec:aitp}). Threads organize work and capability negotiation converts intent into structured, typed payloads suitable for user interfaces, forms, attestations, and payments. When coordination extends to open networks that require escrow and reputation, \textbf{Coral} provides an extension based on DID-backed identity and on-chain reputation systems. MCP or ACP can continue to serve as the internal orchestration layer (\S\ref{subsec:coral}).

\subsection{Is your agent grounded by human web demonstrations?}
The capabilities originating from interactions of the human web are well supported by \textbf{WAP} (\S\ref{subsec:wap}). Browser recordings are compiled into replay lists that form MCP-compatible standalone services. The exact and Smart Replay modes help maintain reliability despite DOM drift while exposing a clear tool interface to agents.

\subsection{Do you need verifiable governance or ethical consensus as part of the interaction?}
Ecosystems that include deliberation, encrypted voting, and auditable outcomes rely on \textbf{LOKA} (\S\ref{subsec:loka}). The protocol formalizes identity, message schemas, and consensus procedures. These mechanisms transform ordinary message exchange into verifiable and ethically responsible decision-making.

\subsection{The overall decision workflow}

\begin{figure}[!ht]
    \centering
    \includegraphics[width=0.8\linewidth]{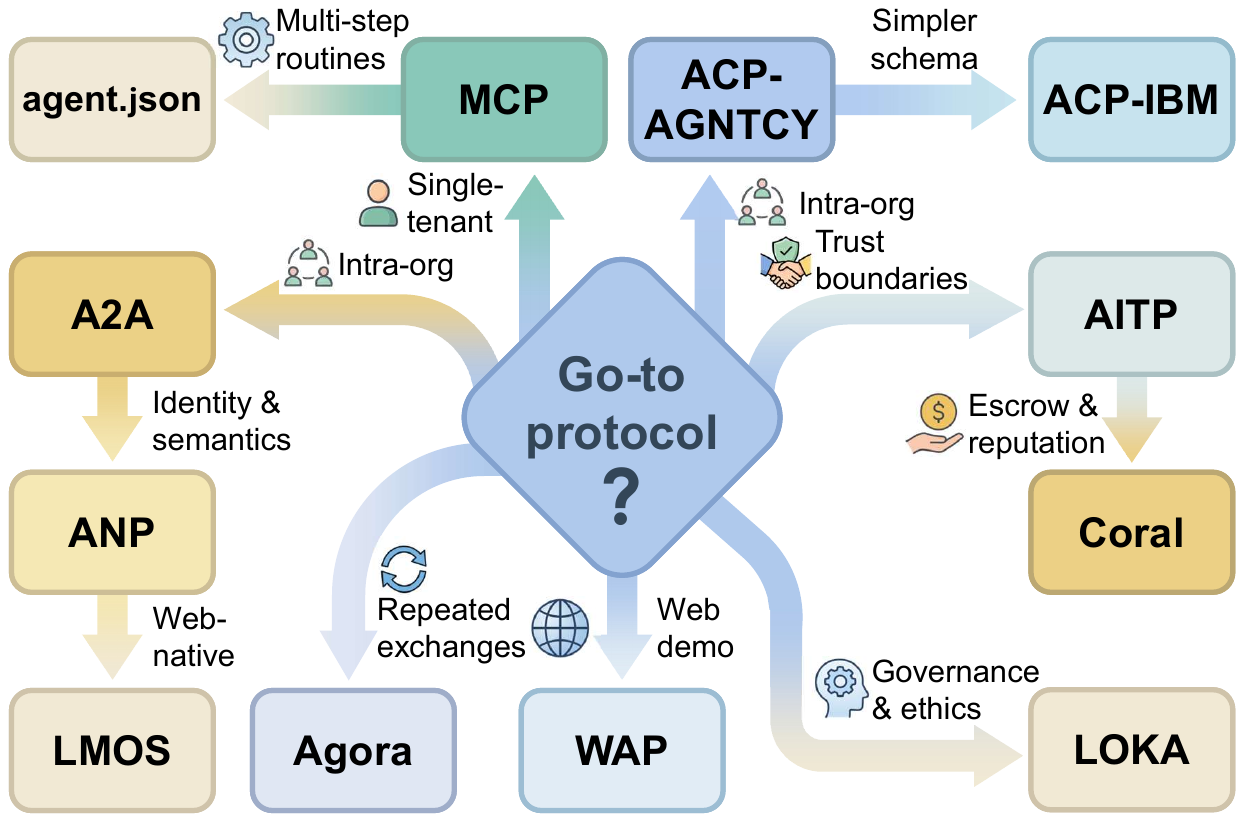}
    \caption{Decision workflow for selecting an agent communication protocol. Starting from deployment and trust-boundary requirements, the workflow routes common application settings to representative protocols, including MCP, ACP-AGNTCY, ACP-IBM, A2A, ANP, LMOS, Agora, Coral, WAP, and LOKA. The goal is to choose the simplest protocol that provides the required communication, syntactic, and semantic guarantees.}
    \label{fig:decision_tree}
\end{figure}
The overall workflow, as shown in Figure~\ref{fig:decision_tree}, can be summarized as follows. Single-tenant, tool-centric deployments converge on \textbf{MCP}, optionally enriched with \textbf{agents.json} (\S\ref{subsec:mcp}, \S\ref{subsec:agents-json}). Intra-organizational orchestration that involves long-running or multi-agent runs prefers \textbf{ACP-AGNTCY} or \textbf{ACP-IBM} (\S\ref{subsec:acp-agntcy}, \S\ref{subsec:acp-ibm}). Cross-organization delegation starts with \textbf{A2A} and extends to \textbf{ANP} or \textbf{LMOS} when identity and linked semantics must persist across domains (\S\ref{subsec:a2a}, \S\ref{subsec:anp}, \S\ref{subsec:lmos}). Repetitive interactions benefit from \textbf{Agora} through stable Protocol Documents and deterministic routines (\S\ref{subsec:agora}). Trust-boundary exchanges adopt \textbf{AITP} and can expand to \textbf{Coral} for escrow and reputation management (\S\ref{subsec:aitp}, \S\ref{subsec:coral}). Demonstration-based capabilities specialize through \textbf{WAP} (\S\ref{subsec:wap}), and governance-driven ecosystems implement \textbf{LOKA} (\S\ref{subsec:loka}). 
From top to bottom, the branches gradually increase the strength of guarantees from transport and syntax toward shared meaning and trust. 
Each protocol helps to reduce session, schema, and semantic drift while adding only the minimal complexity necessary for its context.

\section{Conclusion}
\label{sec:conclusion}

This survey has systematically examined 18 agent communication protocols through a human-inspired, three-layer taxonomy. While our analysis highlights remarkable progress in establishing reliable agent connectivity, it also exposes a fundamental architectural vulnerability within the current ecosystem.

The crux of this vulnerability lies in what we identify as the ``semantic gap.'' Contemporary protocols excel at the communication and syntactic layers, guaranteeing that messages are reliably transmitted and structurally parsed. However, they largely operate under the flawed assumption that successful data transmission equates to shared understanding. Unlike human communication, which naturally employs clarification, confirmation, and contextual repair to navigate ambiguity, most modern agent frameworks strip away these mechanisms for meaning-making. Instead, they push the heavy lifting of intent alignment up to stochastic language models or fragile, application-specific wrappers.

This structural omission is far from a mere theoretical concern; it is the primary driver of the technical debt currently accumulating across multi-agent systems. When protocols lack standardized, native support for verifying meaning, agents become highly susceptible to performative ambiguity and ontological misalignment. A syntactically perfect exchange can easily devolve into an operational breakdown if two agents interpret the same structured payload under different conceptual assumptions. Furthermore, as multi-agent tasks become more complex, the absence of shared contextual memory inevitably leads to contextual drift. The downstream effect is a fragmented landscape where interoperability is artificially maintained through costly integration layers and bespoke adapters, rather than through seamless, protocol-level trust.

To realize the vision of a truly scalable Internet of Agents, the trajectory of protocol design must fundamentally shift. Future standards must look beyond the mechanics of message passing. The community needs to prioritize integrating native semantic grounding---such as explicit clarification loops, verifiable intent schemas, and robust context synchronization---directly into the protocol stack. Only by bridging this semantic gap can we transition from networks of agents that merely exchange data to resilient ecosystems of agents that genuinely understand one another.
\bibliographystyle{acl_natbib}
\bibliography{main.bib}

@inproceedings{hong2023metagpt,
  title={MetaGPT: Meta programming for a multi-agent collaborative framework},
  author={Hong, Sirui and Zhuge, Mingchen and Chen, Jonathan and Zheng, Xiawu and Cheng, Yuheng and Wang, Jinlin and Zhang, Ceyao and Wang, Zili and Yau, Steven Ka Shing and Lin, Zijuan and others},
  booktitle={The twelfth international conference on learning representations},
  year={2023}
}

@inproceedings{wu2024autogen,
  title={Autogen: Enabling next-gen LLM applications via multi-agent conversations},
  author={Wu, Qingyun and Bansal, Gagan and Zhang, Jieyu and Wu, Yiran and Li, Beibin and Zhu, Erkang and Jiang, Li and Zhang, Xiaoyun and Zhang, Shaokun and Liu, Jiale and others},
  booktitle={First conference on language modeling},
  year={2024}
}

@article{schick2023toolformer,
  title={Toolformer: Language models can teach themselves to use tools},
  author={Schick, Timo and Dwivedi-Yu, Jane and Dess{\`\i}, Roberto and Raileanu, Roberta and Lomeli, Maria and Hambro, Eric and Zettlemoyer, Luke and Cancedda, Nicola and Scialom, Thomas},
  journal={Advances in neural information processing systems},
  volume={36},
  pages={68539--68551},
  year={2023}
}

@inproceedings{yao2022react,
  title={React: Synergizing reasoning and acting in language models},
  author={Yao, Shunyu and Zhao, Jeffrey and Yu, Dian and Du, Nan and Shafran, Izhak and Narasimhan, Karthik R and Cao, Yuan},
  booktitle={The eleventh international conference on learning representations},
  year={2022}
}

@article{yang2025survey,
  title={A survey of ai agent protocols},
  author={Yang, Yingxuan and Chai, Huacan and Song, Yuanyi and Qi, Siyuan and Wen, Muning and Li, Ning and Liao, Junwei and Hu, Haoyi and Lin, Jianghao and Chang, Gaowei and others},
  journal={arXiv preprint arXiv:2504.16736},
  year={2025}
}

@article{kong2025survey,
  title={A survey of llm-driven ai agent communication: Protocols, security risks, and defense countermeasures},
  author={Kong, Dezhang and Lin, Shi and Xu, Zhenhua and Wang, Zhebo and Li, Minghao and Li, Yufeng and Zhang, Yilun and Peng, Hujin and Sha, Zeyang and Li, Yuyuan and others},
  journal={arXiv preprint arXiv:2506.19676},
  year={2025}
}

@article{ehtesham2025survey,
  title={A survey of agent interoperability protocols: Model context protocol (mcp), agent communication protocol (acp), agent-to-agent protocol (a2a), and agent network protocol (anp)},
  author={Ehtesham, Abul and Singh, Aditi and Gupta, Gaurav Kumar and Kumar, Saket},
  journal={arXiv preprint arXiv:2505.02279},
  year={2025}
}

@article{deng2025ai,
  title={Ai agents under threat: A survey of key security challenges and future pathways},
  author={Deng, Zehang and Guo, Yongjian and Han, Changzhou and Ma, Wanlun and Xiong, Junwu and Wen, Sheng and Xiang, Yang},
  journal={ACM Computing Surveys},
  volume={57},
  number={7},
  pages={1--36},
  year={2025},
  publisher={ACM New York, NY}
}

@article{ray2025review,
  title={A Review on Agent-to-Agent Protocol: Concept, State-of-the-art, Challenges and Future Directions},
  author={Ray, Partha Pratim},
  journal={Authorea Preprints},
  year={2025},
  publisher={Authorea}
}

@misc{internet_of_agents,
    title = {The Internet of Agents: Building an Open and Interoperable Agentic AI Ecosystem},
    url = {https://www.ciscolive.com/c/dam/r/ciscolive/global-event/docs/2025/pdf/BRKETI-1009.pdf},
    author = {Guillaume de Saint Marc},
    year = {2025}
}

@misc{anthropic_model_2024,
	title = {Model context protocol},
	url = {https://www.anthropic.com/news/model-context-protocol},
	author = {{Anthropic}},
	year = {2024},
}

@misc{alengineerfoundation_agent_2025,
	title = {Agent protocol},
	url = {https://agentprotocol.ai},
	author = {{AlEngineerFoundation}},
	year = {2025},
}

@misc{google_a2a_2025,
	title = {A2a: Agent2agent protocol},
	url = {https://github.com/google/A2A},
	author = {{Google}},
	year = {2025},
}

@misc{samuele_marro_scalable_2024,
	title = {A scalable communication protocol for networks of large language models},
	url = {https://arxiv.org/abs/2410.11905},
	author = {{Samuele Marro} and {Emanuele La Malfa} and {Jesse Wright} and {Guohao Li} and {Nigel Shadbolt} and Wooldridge, Michael and {Philip Torr}},
	year = {2024},
}

@misc{rajesh_ranjan_loka_2025,
	title = {Loka protocol: A decentralized framework for trustworthy and ethical ai agent ecosystems},
	url = {https://arxiv.org/abs/2504.10915},
	author = {{Rajesh Ranjan} and {Shailja Gupta} and {Surya Narayan Singh}},
	year = {2025},
}

@misc{gaowei_chang_anp_2024,
	title = {Anp: Agent network protocol},
	url = {https://www.agent-network-protocol.com},
	author = {{Gaowei Chang}},
	year = {2024},
}

@misc{near_aitp_2025,
	title = {Aitp: Agent interaction \& transaction protocol},
	url = {https://aitp.dev},
	author = {{NEAR}},
	year = {2025},
}

@misc{eclipse_language_2025,
	title = {Language model operating system (lmos)},
	url = {https://eclipse.dev/lmos},
	author = {{Eclipse}},
	year = {2025},
}

@article{ashwin_srinivasan_implementation_2024,
  title={Multi-Turn Human-LLM Interaction Through the Lens of a Two-Way Intelligibility Protocol},
  author={Mestha, Harshvardhan and Bania, Karan and Liu, Sidong and Srinivasan, Ashwin and others},
  journal={arXiv e-prints},
  pages={arXiv--2410},
  year={2024}
}

@misc{ota_tech_ai_web_2025,
	title = {Web Agent Protocol},
	url = {https://github.com/OTA-Tech-AI/web-agent-protocol},
	author = {{OTA Tech AI}},
	year = {2025},
}

@inproceedings{kqml,
  author       = {Timothy W. Finin and
                  Richard Fritzson and
                  Donald P. McKay and
                  Robin McEntire},
  title        = {{KQML} As An Agent Communication Language},
  booktitle    = {Proceedings of the Third International Conference on Information and
                  Knowledge Management (CIKM'94), Gaithersburg, Maryland, USA, November
                  29 - December 2, 1994},
  pages        = {456--463},
  publisher    = {{ACM}},
  year         = {1994},
  url          = {https://doi.org/10.1145/191246.191322},
  doi          = {10.1145/191246.191322},
  bibsource    = {dblp computer science bibliography, https://dblp.org}
}

@article{fipa2002communicative,
  title={Communicative Act Library Specification},
  author={FIPA},
  journal={http://www. fipa. org/specs/fipa00037},
  year={2002}
}

@incollection{smith1988contract,
  title={The contract net protocol: High-level communication and control in a distributed problem solver},
  author={Smith, Reid G},
  booktitle={Readings in distributed artificial intelligence},
  pages={357--366},
  year={1988},
  publisher={Elsevier}
}

@inproceedings{greenwood2005semantic,
  title={Semantic enhancement of a web service integration gateway},
  author={Greenwood, Dominic and Nagy, Jozef and Calisti, Monique},
  booktitle={Proceedings of the Workshop on Service-Oriented Computing and Agent-Based Engineering (SOCABE’2005), Utrecht, The Netherlands},
  year={2005}
}

@article{martin2004owl,
  title={OWL-S: Semantic markup for web services},
  author={Martin, David and Burstein, Mark and Hobbs, Jerry and Lassila, Ora and McDermott, Drew and McIlraith, Sheila and Narayanan, Srini and Paolucci, Massimo and Parsia, Bijan and Payne, Terry and others},
  journal={W3C member submission},
  volume={22},
  number={4},
  year={2004}
}

@misc{agentunion_acp,
  title        = {Agent Communication Protocol (ACP)},
  author       = {{Agent Communication Protocol Contributors}},
  year         = {2025},
  url ={https://agentcommunicationprotocol.dev/introduction/welcome},
}

@article{coral,
  title={Coral protocol: Open infrastructure connecting the internet of agents},
  author={Georgio, Roman J and Forder, Caelum and Deb, Suman and Rahimov, Andri and Carroll, Peter and G{\"u}rcan, {\"O}nder},
  journal={arXiv preprint arXiv:2505.00749},
  year={2025}
}

@article{shannon1948,
  author  = {Claude E. Shannon},
  title   = {A Mathematical Theory of Communication},
  journal = {Bell System Technical Journal},
  volume  = {27},
  number  = {3},
  pages   = {379--423},
  year    = {1948},
  doi     = {10.1002/j.1538-7305.1948.tb01338.x}
}

@incollection{grice1975,
  author    = {H. Paul Grice},
  title     = {Logic and Conversation},
  booktitle = {Syntax and Semantics, Volume 3: Speech Acts},
  editor    = {Peter Cole and Jerry L. Morgan},
  publisher = {Academic Press},
  address   = {New York},
  pages     = {41--58},
  year      = {1975}
}

@book{clark1996,
  author    = {Herbert H. Clark},
  title     = {Using Language},
  publisher = {Cambridge University Press},
  address   = {Cambridge},
  year      = {1996}
}

@book{sperber1995,
  author    = {Dan Sperber and Deirdre Wilson},
  title     = {Relevance: Communication and Cognition},
  edition   = {2},
  publisher = {Blackwell},
  address   = {Oxford},
  year      = {1995}
}

@inproceedings{franklin1996agent,
  title={Is it an Agent, or just a Program?: A Taxonomy for Autonomous Agents},
  author={Franklin, Stan and Graesser, Art},
  booktitle={International workshop on agent theories, architectures, and languages},
  pages={21--35},
  year={1996},
  organization={Springer}
}

@article{jennings2000agent,
  title={On agent-based software engineering},
  author={Jennings, Nicholas R},
  journal={Artificial intelligence},
  volume={117},
  number={2},
  pages={277--296},
  year={2000},
  publisher={Elsevier}
}

@book{wooldridge2009introduction,
  title={An introduction to multiagent systems},
  author={Wooldridge, Michael},
  year={2009},
  publisher={John wiley \& sons}
}

@inproceedings{milojicic1998masif,
  title={MASIF the OMG mobile agent system interoperability facility},
  author={Milojicic, Dejan and Breugst, Markus and Busse, Ingo and Campbell, John and Covaci, Stefan and Friedman, Barry and Kosaka, Kazuya and Lange, Danny and Ono, Kouichi and Oshima, Mitsuru and others},
  booktitle={International Workshop on Mobile Agents},
  pages={50--67},
  year={1998},
  organization={Springer}
}

@inproceedings{curbera2001web,
  title={Web services: Why and how},
  author={Curbera, Francisco and Nagy, William and Weerawarana, Sanjiva},
  booktitle={Workshop on Object-Oriented Web Services-OOPSLA},
  volume={2001},
  year={2001},
  organization={sn}
}

@book{hohpe2004enterprise,
  title={Enterprise integration patterns: Designing, building, and deploying messaging solutions},
  author={Hohpe, Gregor and Woolf, Bobby},
  year={2004},
  publisher={Addison-Wesley Professional}
}

@misc{openai_function_calling,
  author       = {{OpenAI}},
  title        = {Function calling},
  url={https://developers.openai.com/api/docs/guides/function-calling},
  note         = {OpenAI API documentation. Accessed: 2026-03},
  year         = {2025}
}

@misc{chase2022langchain,
  author       = {Harrison Chase},
  title        = {LangChain},
  url={https://github.com/langchain-ai/langchain},
  note         = {GitHub repository. Accessed: 2026-03},
  year         = {2022}
}

@misc{liu2023llamaindex,
  author       = {Jerry Liu and others},
  title        = {LlamaIndex},
  url={https://github.com/run-llama/llama_index},
  note         = {GitHub repository. Accessed: 2026-03},
  year         = {2023}
}

@misc{openai_plugins,
  author = {{OpenAI}},
  title  = {Plugins},
  url    = {https://developers.openai.com/codex/plugins/},
  note   = {OpenAI Developers documentation. Accessed: 2026-03},
  year   = {2026}
}

@misc{autogpt,
  author = {{Significant Gravitas}},
  title  = {AutoGPT: Build, Deploy, and Run AI Agents},
  url    = {https://github.com/Significant-Gravitas/AutoGPT},
  note   = {GitHub repository. Accessed: 2026-03},
  year   = {2023}
}

@misc{crewai,
  author = {{crewAIInc}},
  title  = {CrewAI: Fast and Flexible Multi-Agent Automation Framework},
  url    = {https://github.com/crewAIInc/crewAI},
  note   = {GitHub repository. Accessed: 2026-03},
  year   = {2024}
}

@misc{ag2,
  author = {{AG2 Contributors}},
  title  = {AG2: Open-Source AgentOS for AI Agents},
  url    = {https://github.com/ag2ai/ag2},
  note   = {GitHub repository. Accessed: 2026-03},
  year   = {2025}
}

@article{milev2025toolfuzz,
  title={ToolFuzz--Automated Agent Tool Testing},
  author={Milev, Ivan and Balunovi{\'c}, Mislav and Baader, Maximilian and Vechev, Martin},
  journal={arXiv preprint arXiv:2503.04479},
  year={2025}
}

@techreport{sheriff2024metadata_manifest,
  author      = {Akram Sheriff},
  title       = {Dynamic LLM Agent Metadata Manifest-Based Discovery of Agents in an LLM Agentic Application Platform},
  institution = {Technical Disclosure Commons},
  number      = {7522},
  year        = {2024},
  month       = nov,
  url         = {https://www.tdcommons.org/dpubs_series/7522},
  note        = {Accessed: 2026-03}
}

\end{document}